\documentclass[
  journal=pasa,
  manuscript=research-paper, 
  year=2025,
  volume=xx,
]{cup-journal}

\usepackage{aas-macros}
\usepackage{graphicx}
\usepackage{xcolor}
\usepackage{microtype,siunitx}
\usepackage{amssymb}
\usepackage{amsmath}
\usepackage{hyperref}

\hypersetup{
    colorlinks,
    citecolor=blue,
    linkcolor=blue,
    urlcolor=blue,
}

\usepackage{enumitem, xcolor}
  
\usepackage{booktabs} 
\usepackage{threeparttable,longtable}  
\usepackage{makecell}

\usepackage[skip=0.5ex]{subcaption}
\usepackage{cprotect}
\usepackage{multirow}

\definecolor{dodgerblue}{RGB}{30, 144, 255}

\DeclareUnicodeCharacter{2009}{\,}

\title{Long-term spin-down and low luminosity regime in the Be/X-ray
binary pulsar GX 304-1}

\author{Amar Deo Chandra}
\affiliation{Aryabhatta Research Institute of Observational Sciences, Manora Peak, Nainital, Uttrakhand, 263001, India}
\email[A. D. Chandra]{amardeo@aries.res.in, amar.deo.chandra@gmail.com}

\doi{10.1017/pasa.20XX.XX}

\received {dd Mmm YYYY}
\revised  {dd Mmm YYYY}
\accepted {dd Mmm YYYY}
\published{dd Mmm YYYY}


\begin{document}

\begin{abstract}
We carry out timing and spectral studies of the Be/X-ray binary pulsar GX 304-1 using \textit{NuStar} and \textit{XMM-Newton} observations. We construct the  long-term spin period evolution of the pulsar which changes from a long-term spin-up ($\sim 1.3 \times 10^{-13}$\,Hz\,s$^{-1}$) to a long-term spin-down ($\sim -3.4 \times 10^{-14}$\,Hz\,s$^{-1}$) trend during a low luminosity state ($\sim 10^{34-35}$\,erg\,s$^{-1}$).
A prolonged low luminosity regime ($L_X \sim 10^{34-35}$\,erg\,s$^{-1}$) was detected during 2005-2010 and spanning nearly five years since 2018 December. The \textit{XMM-Newton} and \textit{NuStar} spectra can be described with a power law plus blackbody model having an estimated luminosity of $\sim 2.5 \times 10^{33}$\,erg\,s$^{-1}$ and $\sim 3.6 \times 10^{33}$\,erg\,s$^{-1}$ respectively. The inferred radius of the blackbody emission is about 100-110 m which suggests a polar-cap origin of this component. From long-term ultraviolet observations of the companion star, an increase in the ultraviolet signatures is detected preceding the X-ray outbursts. The spectral energy distribution of the companion star is constructed which provides a clue of possible UV excess when X-ray outbursts were detected
from the neutron star compared to the quiescent phase. We explore plausible mechanisms to explain the long-term spin-down and extended low luminosity manifestation in this pulsar. We find that sustained accretion from a cold disc may explain the prolonged
low luminosity state of the pulsar since December 2018 but the pulsar was undergoing normal accretion during the low luminosity period spanning 2005-2010.\\
\end{abstract}

\section{Introduction}
\label{sec:intro}

The Be/X-ray binary (BeXRB) pulsar GX 304-1 was first detected using balloon-borne observations in 1967 \citep{mcclintock1971rapid}. It was subsequently detected using \textit{Uhuru} observations and listed in the \textit{Uhuru} catalogue \citep{giacconi1974third}. X-ray pulsations having a periodicity of about 272 s were detected from this source using \textit{SAS-3} observations \citep{mcclintock1977discovery}. Similar pulsation period was inferred using \textit{Ariel V} observations of the source \citep{huckle1977discovery}. Spectral investigations of the pulsar were performed using follow-up observations \citep{maurer1982balloon,white1983accretion} and the spectrum was found to be hard ($\Gamma \sim$ 2) and described using an absorbed power law. The orbital period of the binary was estimated to be about 132.5 d from modulations in the X-ray outbursts from \textit{Vela 5B} observations \citep{priedhorsky1983long}.  The optical companion of the binary pulsar was identified by \cite{bradt1977optical}. The companion star V850 Cen was found to be of B2Vne type using optical observations \citep{mason1978optical,thomas1979lambda,parkes1980shell} and photometric studies of the companion star were carried out \citep{menzies1981photoelectric}. The distance of the source was estimated to be 2.4$\pm$0.5 kpc by \citet{parkes1980shell} and recent \textit{Gaia} observations pin down the distance to 2.01$\pm$0.15 kpc \citep{treuz2018distances}. The pulsar was not detected using \textit{EXOSAT} monitoring observations in 1984 July/August which suggested that the source was in an \textquotedblleft off\textquotedblright ~state \citep{pietsch1986exosat}. The putative cause of this intriguing \textquotedblleft off\textquotedblright ~state was attributed to the loss in the decretion disc around the companion star from long-term optical observations \citep{pietsch1986exosat,corbet1986long}.

The pulsar lay dormant for about three decades until 2008 when it was detected using \textit{INTEGRAL} \citep{manousakis2008integral} and regular outbursts spaced by the orbital period were detected until mid-2013 \citep{yamamoto2009maxi,mihara2010maxi,krimm2010swift,nakajima2010maxi,kuhnel2010recent,yamamoto2011maxia,yamamoto2011maxib,yamamoto2012maxi}. The orbital period was refined to 132.1885$\pm$0.022 d using \textit{MAXI} observations of the recurrent outbursts \citep{sugizaki2015luminosity}. The pulsation period detected during the 2010 August period was about 275.4 s  which suggested that the pulsar had spun down by about 3 s during the quiescent period lasting for about 28 yr \citep{devasia2011timing,yamamoto2011discovery}. A cyclotron
absorption feature at around 54 keV was detected in the spectrum using \textit{RXTE} observations during the 2010 August outburst \citep{yamamoto2011discovery} which was confirmed using \textit{INTEGRAL} observations of the source \citep{klochkov2012outburst}. The pulsar remained quiescent until a weak X-ray brightening was detected in
2016 January \citep{nakajima2016maxia} and then in 2016 May \citep{nakajima2016maxib,sguera2016integral,rouco2016swift}. Thereafter, the pulsar lay in a low luminosity state and accretion-induced pulsations were detected from the pulsar using \textit{NuStar} observations of the source during a low luminosity state in 2018 June which was suggested to occur from a cold disc \citep{escorial2018discovery}. The pulsar lay in an enigmatic low luminosity state for almost a year from around 2017 September, wherein the \textit{Swift}/X-ray Telescope (XRT) count rate varied by a factor of only $\sim$2-3 and was not tied to any particular orbital phase \citep{escorial2018discovery}. This peculiar behaviour was not well understood and was surmised to be due to accretion from a cold disc \citep{escorial2018discovery}.

In this paper, we investigate the long-term spin period evolution and low luminosity regime of GX 304-1 using multiwavelength observations.

The paper is organized as follows. We describe observations from the \textit{NuStar}, the \textit{XMM-Newton} and the \textit{Swift} missions and their data analysis procedures in Section 2. In section 3, we carry out timing and spectral studies using \textit{NuStar} and \textit{XMM-Newton} observations and construct the long-term spin evolution of the pulsar.
This is followed by exploring the X-ray activity of the pulsar during the period when the spin evolution of the pulsar was surmised to have changed from a long-term spin-up to a long-term spin-down trend. The long-term photometric and ultraviolet observations of the companion star are explored. Thereafter, we study the long-term low-level X-ray activity of the pulsar
using archival \textit{Neil Gehrels Swift Observatory} observations. In Section 4, we discuss possible mechanisms that can lead to a long-term spin-down in the pulsar and sustained low luminosity manifestation of the pulsar. In addition, we also probe the origin of soft X-ray excess in the low luminosity spectra of the neutron star and explore the spectral energy distribution of the companion star.  We summarise our findings in Section 5.

\section{Observations and data reduction}
We analyse unpublished \textit{NuStar} and \textit{XMM-Newton} observations of GX 304-1 from 2022 January and 2023 July respectively. The source was observed on several occasions during regular outbursts and quiescent phases using the \textit{Neil Gehrels Swift Observatory} since 2005 April. The X-ray activity of the pulsar during the period 2012 January 1 until 2018 October 30 was explored by \cite{escorial2018discovery} using the X-ray Telescope (XRT) onboard the \textit{Neil Gehrels Swift Observatory}. We explore the X-ray activity of the pulsar using unpublished \textit{Swift}/XRT observations from 2005 April 6-2010 March 15 and 2018 December 17-2024 February 27.

\subsection{\textit{NuStar} observation}
\begin{figure}
\centering
  \includegraphics[width=\columnwidth]{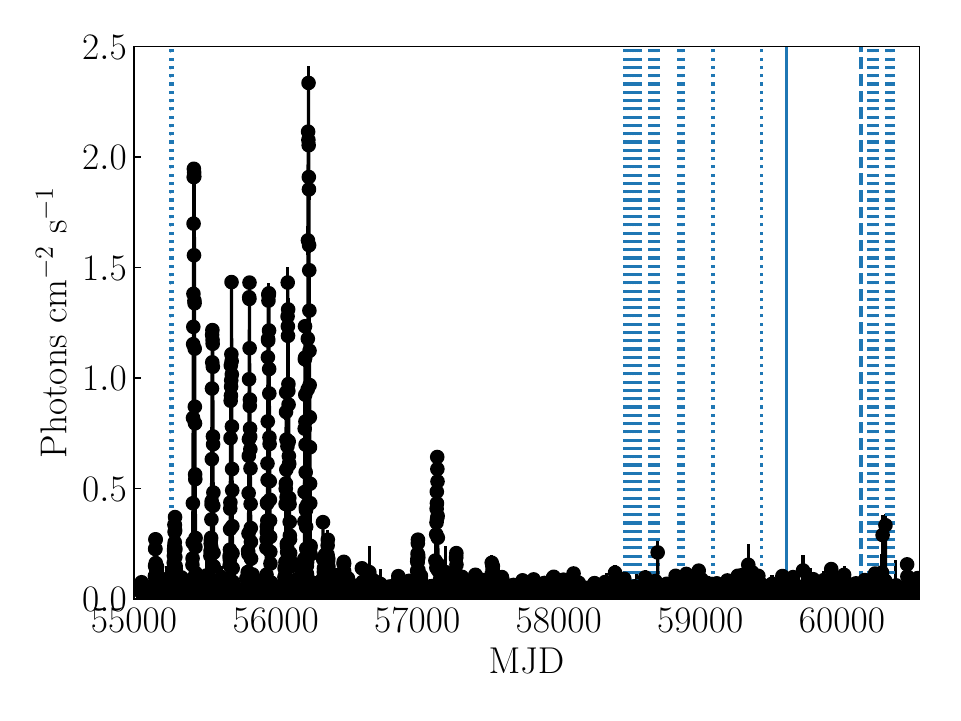}
  \caption{\textit{MAXI} one day averaged light curve of GX 304-1 in
the 2-20 keV energy band spanning the duration MJD 55054.5 (2009 August 11) until MJD 60545 (2024 August 23). The solid, dotted and dashed vertical lines indicate the epochs of \textit{NuStar}, \textit{Swift} and \textit{XMM-Newton} observations respectively. A few epochs of \textit{Swift} observations before 2009 August 11 are not shown here as they precede the time since the \textit{MAXI} mission became operational.}
 \label{f00}
\end{figure}

The  \textit{Nuclear Spectroscopic Telescope Array} (\textit{NuStar}) is a hard X-ray telescope consisting of two identical modules (FPMA and FPMB) operating in the energy range 3-79 keV
\citep{harrison2013nuclear}. \textit{NuStar} observed GX 304-1 on 2022 January 29 (MJD 59608.5, ObsID 30701015002) for a total duration of about 174 ks. Figure \ref{f00} shows the one-day averaged monitoring observations of GX 304-1 from the \textit{MAXI} mission in the 2-20 keV energy band and the epoch of the \textit{NuStar}
observation is marked with a vertical solid line. The orbital phase of the \textit{NuStar} observation was about 0.65 (using the orbital
parameters $P_{\rm{orb}}$=132.189 d and $T_0$=MJD 55425.6 \citep{sugizaki2015luminosity}). We have extracted data separately from both the modules using the standard \textit{NuStar} data analysis software (\textsc{nustardas v2.1.4}) included in \textsc{heasoft v6.34}\footnote{\url{https://heasarc.gsfc.nasa.gov/docs/software/lheasoft/}} along with the calibration database \textsc{caldb} version 20240826. The \textsc{nupipeline} task (version 0.4.12) was run with \textsc{saamode}=strict and \textsc{tentacle}=yes because the background event rates were high to obtain clean event files.

The event files were barycentered using the
\textsc{ftools} task  ``barycorr''. We used a circular region of radius $80''$ centred on the source to extract the source events for both FPMA and FPMB modules. The \textsc{nuproducts} task was used to generate light curves binned in 1 s.

\subsection{\textit{XMM-Newton} observation}
\textit{XMM-Newton} \citep{jansen2001xmm} observations of GX 304-1 were carried out on 2023 July 11 (MJD 60137, ObsID 0931790601) for a total duration of about 8 ks at an orbital phase of about 0.65 (using the orbital
parameters $P_{\rm{orb}}$=132.189 d and $T_0$=MJD 55425.6 \citep{sugizaki2015luminosity}) which is similar to the orbital phase of the \textit{NuStar} 2022 observation. \textit{XMM-Newton} has three European Photon Imaging Camera (EPIC) cameras viz. one pn and two Metal Oxide Semi-conductor (MOS) cameras \citep{struder2001european,turner2001european}. The EPIC cameras were operated in the large window mode during this observation. The time resolution for the pn camera and the two MOS cameras used during this observation was 48 ms and 0.9 s respectively. Observation Data Files (ODFs) were processed using version 21.0.0 of the \textit{XMM-Newton} Science Analysis System (\textsc{sas})\footnote{\url{https://www.cosmos.esa.int/web/xmm-newton/sas}} and the Current Calibration Files (CCF)\footnote{\url{https://www.cosmos.esa.int/web/xmm-newton/current-calibration-files}} released on 2024 April 29. We searched for possible intervals of high instrumental background which yielded a negative result. The effective source exposures were about 5 ks and 7 ks for the pn and MOS cameras respectively. We corrected the event arrival times to the solar system barycenter using \textsc{sas}'s ``barycen'' task. A circular region of radius $30''$ and $40''$ was used to extract the source events for the pn and the MOS cameras. Background events were extracted using circular regions away from the source with radii of $70''$, $60''$, and $50''$ for the pn, MOS1, and MOS2 cameras, respectively. We selected all the events in the energy range 0.3-10 keV, with pattern range 0-4 for the pn camera and 0-12 for the two MOS cameras. The averaged count rate for the pn and MOS cameras was about 1 $\rm{counts ~s^{-1}}$ and 0.4 $\rm{counts ~s^{-1}}$ respectively. The background contribution to the total count rate was negligible for EPIC cameras (averaged count rate of about 0.01 $\rm{counts ~s^{-1}}$).

\subsection{\textit{Swift} observations}
The \textit{Neil Gehrels Swift Observatory} \citep{gehrels2004swift} observed GX 304-1 for various durations spanning the period from 2005 April
6 (MJD 53466) until 2024 February 27 (MJD 60367.8). \textit{Swift} has three onboard instruments viz. the Burst Alert Telescope (BAT, \cite{krimm2013swift}), the X-Ray Telescope (XRT, \cite{burrows2005swift}) and the Ultraviolet/Optical Telescope (UVOT, \cite{roming2005swift}). We have used the data from only the XRT and the UVOT instruments in this work. The log of \textit{Swift}/XRT  observations used in this study before regular outbursts were detected from the pulsar around 2010 April is given in Table {\ref{tab1}}. The pulsar entered a low luminosity state around 2017 September and was found to be in the same state until around 2018 October \citep{escorial2018discovery}. We have analysed all available \textit{Swift} observations since 2018 October to probe if the pulsar continues to remain in a similar low accretion regime and these observations are listed in Table {\ref{tab2}}. \textit{Swift}/XRT observations were mostly carried out in the photon counting (PC) mode having a time resolution of about 2.5 s. The  \textit{Swift}/XRT light curves for each epoch were extracted using the online tools \citep{evans2009methods}
\footnote{\url{https://www.swift.ac.uk/user_objects/}}
hosted by the UK Swift Science Data Centre. The typical on-source effective exposures are about 1 ks for each \textit{Swift}/XRT pointing.

\textit{Swift}/UVOT archival observations of GX 304-1 are available from 2006 September 20 (MJD 53998.1)
until 2024 February 27 (MJD 60367.8), having typical exposure times of about a few tens of seconds to a few hundreds of seconds at different epochs. UVOT observations were performed with the filters V ($\lambda$=546.8 nm, $\delta{\lambda}$=76.9 nm), B ($\lambda$=439.2 nm, $\delta{\lambda}$=97.5 nm), U ($\lambda$=346.5 nm, $\delta{\lambda}$=78.5 nm), UVW1 ($\lambda$=260.0 nm, $\delta{\lambda}$=69.3 nm), UVM2 ($\lambda$=224.6 nm, $\delta{\lambda}$=49.8 nm) and UVW2 ($\lambda$=192.8 nm, $\delta{\lambda}$=65.7 nm)\footnote{\url{https://www.mssl.ucl.ac.uk/www_astro/uvot/uvot_instrument/filterwheel/filterwheel.html}}. The Level2 UVOT  data were analysed using the \textsc{uvotsource} tool from \textsc{heasoft} v6.34 to determine magnitudes and fluxes. A circular region of radius $5''$ and $20''$ were used to select the source region and background region respectively.
The source was detected in the UVOT filters V, B, U, UVW1, UVM2 and UVW2 having an averaged magnitude of about 14, 15.7, 16.6, 17.9, 20.9 and 19.2 respectively.

\newpage
\section{Analysis and results}

\subsection{Timing analysis}
\begin{figure}
\centering
  \includegraphics[width=\columnwidth]{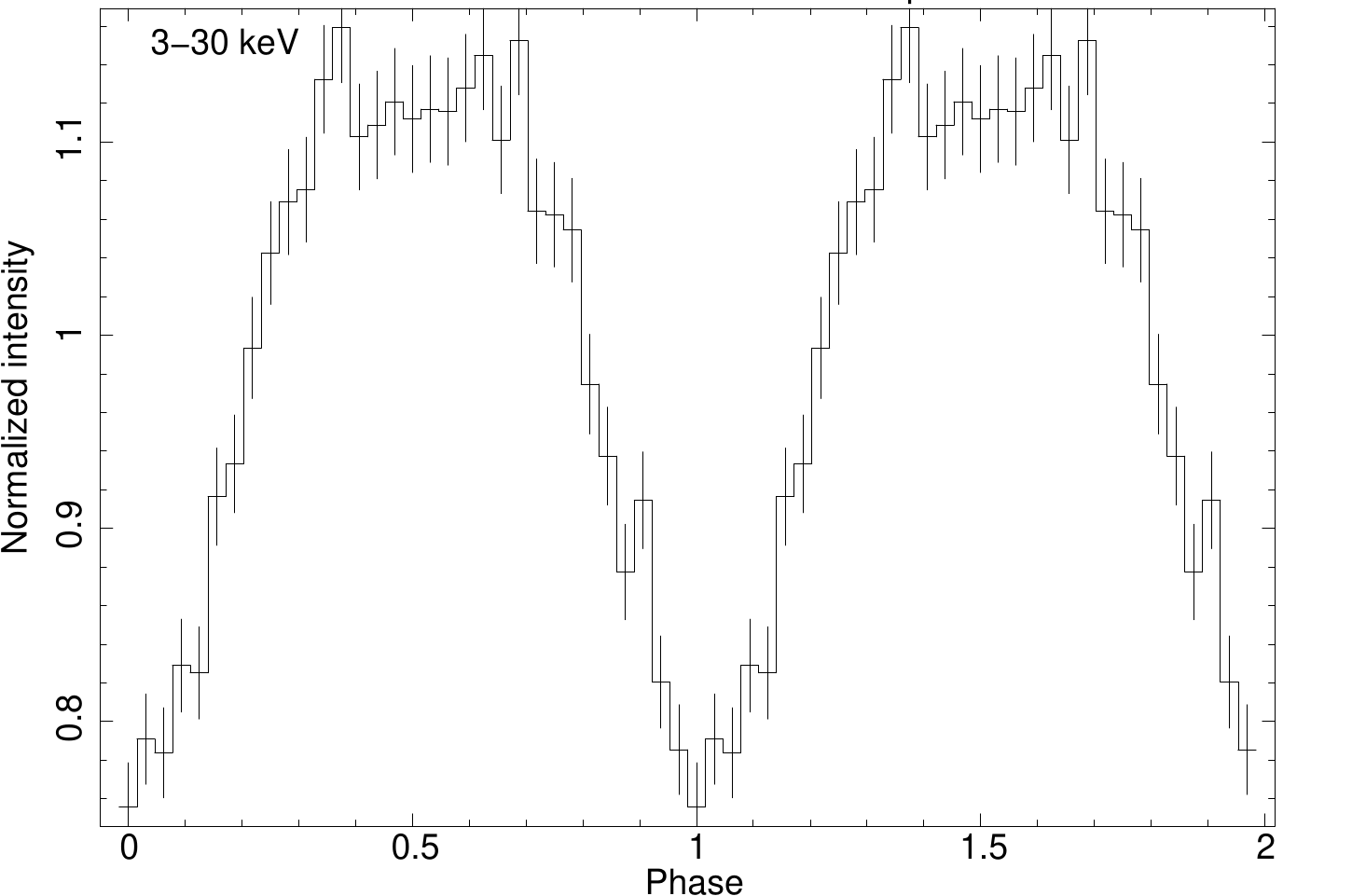}
  \caption{\textit{NuStar} folded profile in the energy band of 3-30 keV obtained from the combined FPMA and FPMB light curve.} 
 \label{f01a}
\end{figure}

 We used the \textit{NuStar} combined FPMA and FPMB light curves in the 3-30 keV energy band for timing analysis. The FTOOLS\footnote{\url{https://heasarc.gsfc.nasa.gov/ftools/ftools_menu.html}} subroutine \textsc{efsearch} was used to search for pulsations using the chi-squared maximization method \citep{leahy1983searches}. The inferred spin period from \textit{NuStar} 2022 January (MJD 59608.5) observations is 275.425$\pm$0.003 s. The error on the measured spin period is estimated by fitting a Gaussian to the chi-square versus spin period plot obtained by \textsc{efsearch}. The 1-$\sigma$ error on the Gaussian centre estimate is taken as the error on the spin period.
The estimated spin period is consistent with the known spin period of GX 304-1. The previous measured spin period of the source was about 275.12 s on 2018 June 3 (MJD 58272.25) \citep{escorial2018discovery}, which suggests that the source has spun down by about 0.3 s.
The folded pulse profile in the 3-30 keV energy band using \textit{NuStar} observations is shown in Figure \ref{f01a}.

The estimated pulsed fraction ($\mathrm{PF}=(\mathrm{I}_{\mathrm{max}}-\mathrm{I}_{\mathrm{min}})/(\mathrm{I}_{\mathrm{max}}+\mathrm{I}_{\mathrm{min}})$, where $\mathrm{I}_{\mathrm{max}}$
and $\mathrm{I}_{\mathrm{min}}$ are the maximum and minimum intensities in the
folded proﬁle respectively) in the 3-30 keV energy band is about 21 per cent. A similar folded pulse profile and PF in the 3-30 keV energy band was obtained from the  \textit{NuStar} observations of the pulsar carried on 2018 June 3 \citep{escorial2018discovery} which suggests that the profile shape has not changed during this period. Similar PF of about 20 per cent in the 3-20 keV energy band was obtained during the February-January 2012 INTEGRAL observations of the pulsar \citep{klochkov2012outburst}.

\begin{figure}
\centering
  \includegraphics[width=\columnwidth]{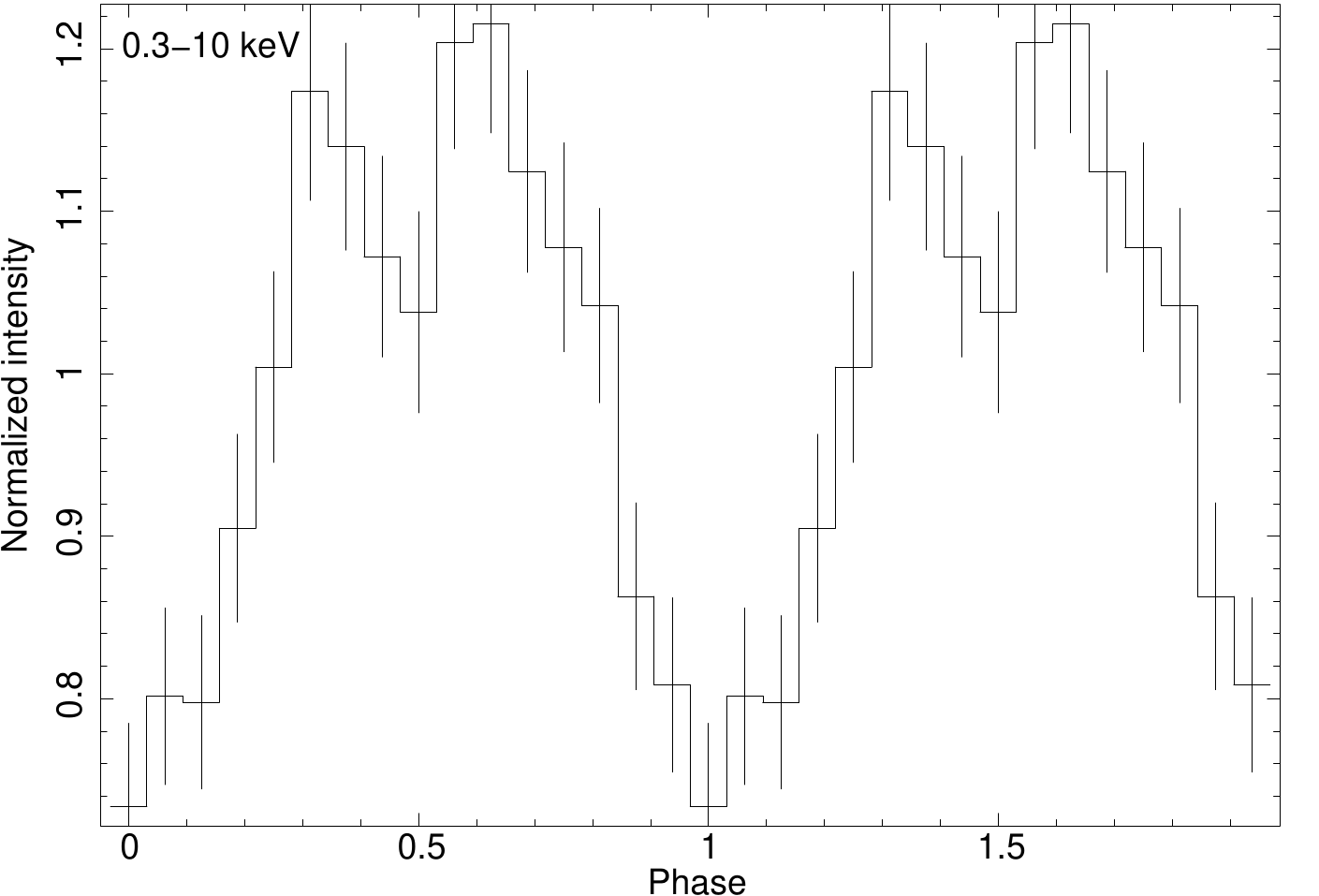}
  \caption{\textit{XMM-Newton} folded profile of GX 304-1 in the energy band of 0.3-10 keV using data from EPIC pn observations.} 
 \label{f01b}
\end{figure}

We have used the \textit{XMM-Newton} pn data for timing analysis as the pulsar is relatively fainter in the MOS data. We used the pn 0.3-10 keV light curve binned in 10 s for timing analysis. The spin period estimated using \textsc{efsearch} is 275.5$\pm$0.3 s. The error on the estimated spin period is obtained by the method described earlier.
The folded profile in the 0.3-10 keV band using the pn light curve is shown in Figure \ref{f01b} which shows the presence of two peaks at low energies. Similar folded pulse profiles having two dominant peaks in soft X-rays (< 10 keV) have been detected during outbursts of the pulsar in 2010 August \citep{devasia2011timing}. The estimated PF from EPIC pn observations of the pulsar is about 25 per cent.

\subsection{Spectral analysis}
\begin{figure}
\centering
  \includegraphics[width=\columnwidth]{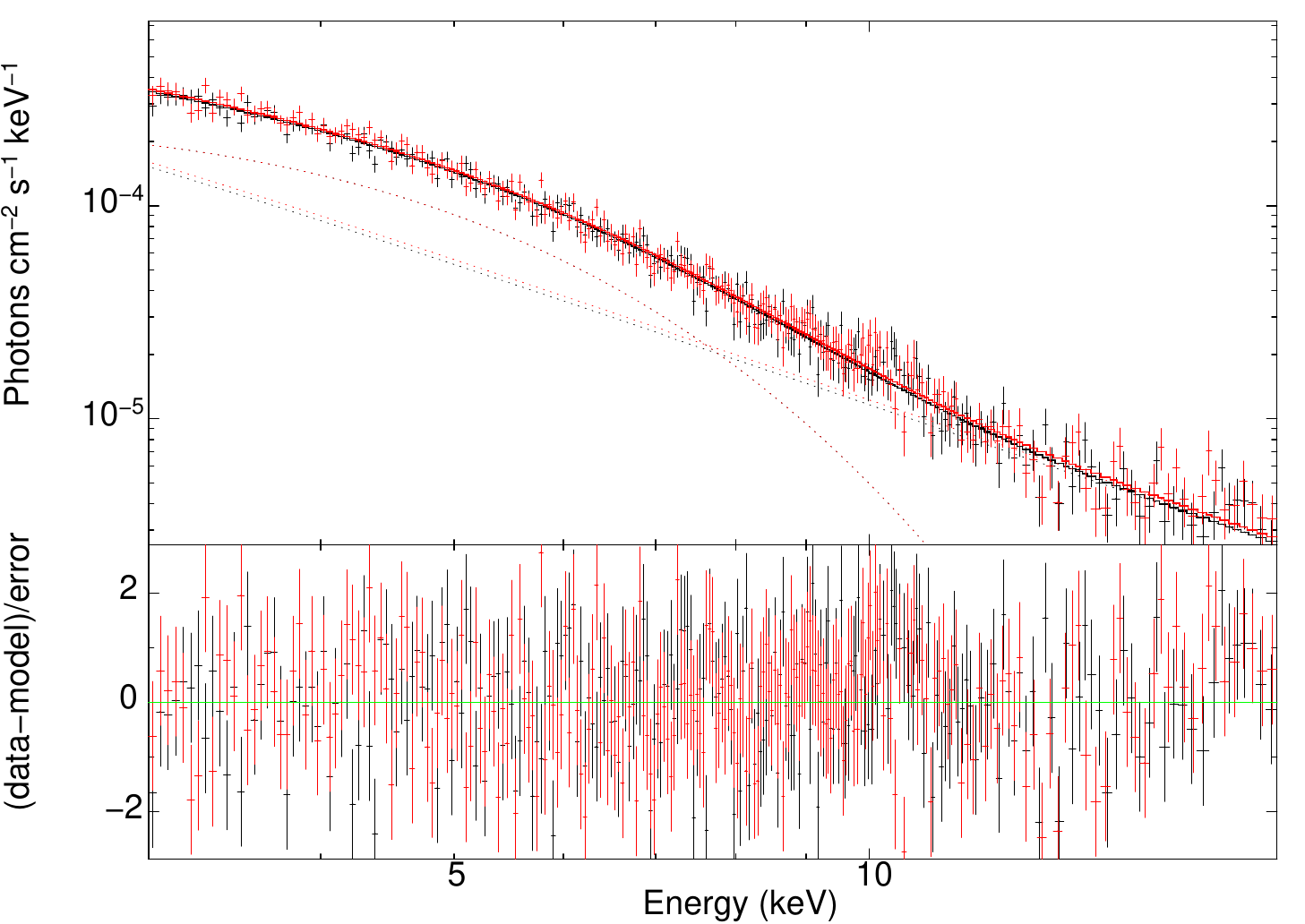}
  \caption{Spectrum of GX 304-1 obtained from \textit{NuStar}’s FPMA (black) and FPMB (red), along with the best-fitting \textit{PL+BB} model. The residuals between the data and the model are shown in the lower panel.}
 \label{f03}
\end{figure}

The \textit{NuStar} spectra were extracted from the same extraction parameters used for the light curves. We have done a combined fit of \textit{NuStar}’s FPMA and FPMB data using \textsc{xspec} 12.14.1. The data have been binned using a minimum of 30 counts per bin. We have used a multiplicative model constant \textsc{const} to account for cross-instrument calibration uncertainties. The value of this constant was frozen at 1 for FPMA and was kept free for FPMB. The spectral fitting using \textsc{xspec} 12.14.1 \citep{arnaud1996astronomical} was limited to 3-20 keV as the background dominates at higher energies. We fitted the combined spectra using an absorbed \textit{PL+BB} model as shown in Figure \ref{f03}. We have used the \textit{tbabs} model \citep{wilms2000absorption} to take care of absorption in the spectrum during spectral fitting. As keeping the $N_{\rm H}$ free did not allow us to constrain the absorption, we kept it fixed at $9.6 \times 10^{21}$ cm$^{-2}$ which is the interstellar Galactic absorption along the direction of this source \citep{bekhti2016hi4pi} \footnote{\url{https://heasarc.gsfc.nasa.gov/cgi-bin/Tools/w3nh/w3nh.pl}}. The best-fit parameters obtained using this model were $\Gamma = 2.21^{+0.12}_{-0.14}$, $kT_{\rm BB} = 1.17^{+0.03}_{-0.02}$ keV, with $\chi^{2}_{\nu}$/d.o.f = 1.02/476. Using a source distance of 2.01 kpc, we obtained a \textit{BB} radius $R_{\rm BB}= 100^{+7}_{-8} $ m. The unabsorbed flux in the energy range 3-20 keV
 is $\sim 7.5 \pm 0.1 \times10^{-12}$ erg cm$^{-2}$ s$^{-1}$ , which implies a source luminosity of about $3.6 \times10^{33}$ erg s$^{-1}$ for a distance of 2.01 kpc. The estimated luminosity in the 0.5-100 keV energy range is $\sim 1 \times10^{34}$ erg s$^{-1}$ using the WebPIMMS tool\footnote{\url{https://heasarc.gsfc.nasa.gov/cgi-bin/Tools/w3pimms/w3pimms.pl}}.

The \textit{XMM-Newton} spectra for the pn and MOS cameras were extracted using the same extraction parameters used to extract the source and background light curves. The response matrices and ancillary files were generated using the tasks \texttt{rmfgen} and \texttt{arfgen}. The pn and MOS spectra were rebinned with a minimum of 25 counts per bin to ensure the applicability of the  $\chi^{2}$ statistics.
We have used \textsc{xspec} 12.14.1 \citep{arnaud1996astronomical} for spectral fitting in the energy band of 0.5-10 keV. All spectral uncertainties are given at the 90\%
confidence level for each parameter.
We checked that the separate fits of the pn and the MOS data gave similar results and then fitted them simultaneously to improve the statistics. The inter-calibration among the three instruments was accounted for using \textsc{const}.

\begin{figure*}[!htbp]
        \begin{subfigure}[b]{0.48\textwidth}
                \includegraphics[width=\linewidth,keepaspectratio=true]
                {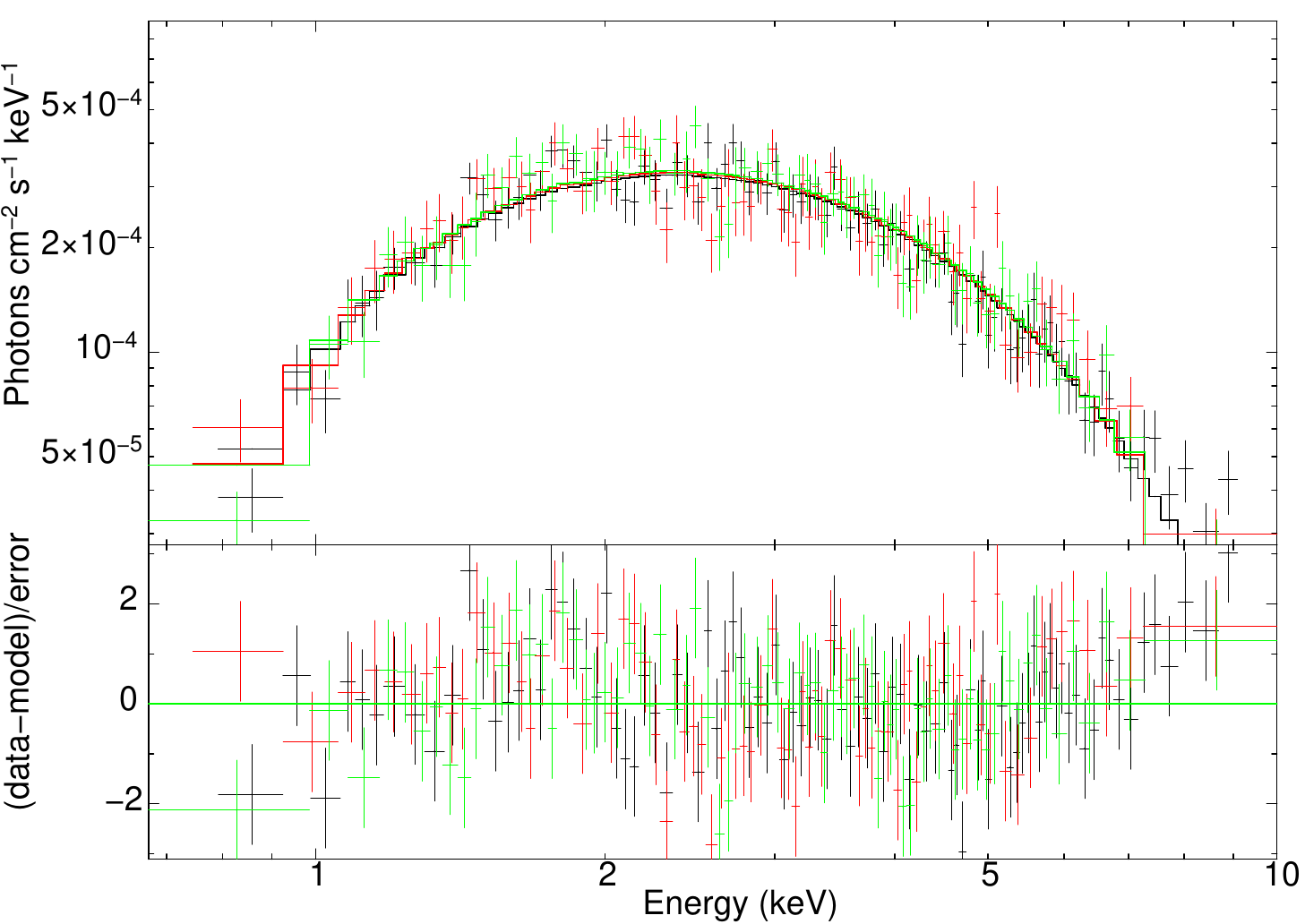}
                \caption{}
        \end{subfigure}
        \begin{subfigure}[b]{0.48\textwidth}
                \includegraphics[width=\linewidth]{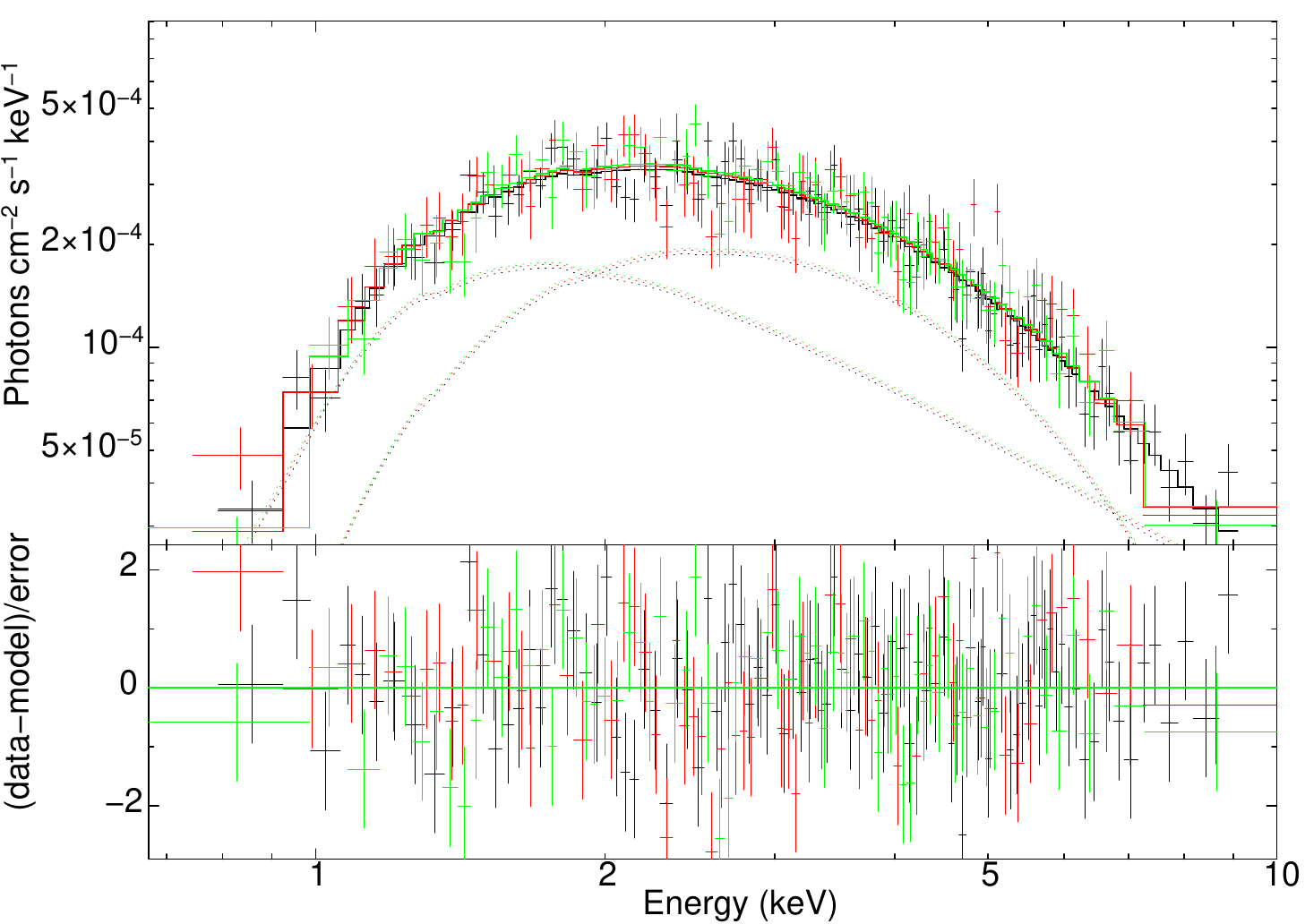}
                \caption{}
        \end{subfigure}
        \caption{(i) Spectrum of GX 304-1 fitted with the
absorbed blackbody model. The spectra of the pn, MOS1 and MOS2 cameras are
shown in black, red, and green, respectively. The residuals between the data and the model are shown in the lower panel. (ii) The same spectra shown in (i) fitted with the absorbed powerlaw+blackbody model with the residuals between the data and the model shown in the lower panel.}
        \label{f02}
\end{figure*}

We first fitted the spectra using an absorbed power-law (\textit{PL}) model. We obtained hydrogen column density $N_{\rm H} = (2.36\pm0.14)\times 10^{22}$ cm$^{-2}$ and a photon-index $\Gamma$ = 1.83$\pm$0.06, with $\chi^{2}_{\nu}$/d.o.f. = 1.47/249. The fit using the absorbed-PL model was unacceptable and so we fitted the spectra using an absorbed blackbody (\textit{BB}) model (Figure \ref{f02}(i)) which resulted in $N_{\rm H} = (7.14\pm0.06)\times 10^{21}$ cm$^{-2}$, a \textit{BB} temperature $kT_{\rm BB} = 1.16 \pm 0.02$ keV,  with $\chi^{2}_{\nu}$/d.o.f. = 1.12/245. It should be noted that the fitted $N_{\rm H}$ in this model is slightly less than the interstellar absorption ($N_{\rm H} \sim 9.6 \times 10^{21}$ cm$^{-2}$) along the direction of this source.  Using a source distance of 2.01 kpc, we obtained a \textit{BB} radius $R_{\rm BB}= 132\pm 5 $ m. The
unabsorbed flux is $8.0 \pm 0.2 \times10^{-12}$ erg cm$^{-2}$ s$^{-1}$  in the energy range of 0.5-10 keV, which implies a source luminosity of $3.9 \pm 0.1 \times10^{33}$ erg s$^{-1}$ for a distance of 2.01 kpc.

We also tried fitting the \textit{PL+BB} model as shown in Figure \ref{f02}(ii) and the best-fit parameters obtained using this model were $N_{\rm H} = (1.48^{+0.29}_{-0.31}) \times 10^{22}$ cm$^{-2}$, $\Gamma = 1.65^{+0.44}_{-0.43}$, $kT_{\rm BB} = 1.10^{+0.10}_{-0.09}$ keV, with $\chi^{2}_{\nu}$/d.o.f = 0.91/243. Using the F-test analysis, the probability that the improvement in the fit
occurs by chance was found to be $1\times 10^{-24}$ and $1\times 10^{-11}$ in comparison with the single PL and BB models respectively. Using a source distance of 2.01 kpc, we obtained a \textit{BB} radius $R_{\rm BB}= 113^{+21}_{-15} $ m.
The unabsorbed flux
in the energy range 0.5-10 keV
 is $5.2 \pm 1.3 \times10^{-12}$ erg cm$^{-2}$ s$^{-1}$ , which implies a source luminosity of about $2.5 \pm 0.6 \times10^{33}$ erg s$^{-1}$ for a distance of 2.01 kpc. The estimated luminosity in the 0.5-100 keV energy range is $\sim 7 \times10^{33}$ erg s$^{-1}$ using the WebPIMMS tool.

\subsection{Long-term spin evolution in GX 304-1}

\begin{figure}
\centering
  \includegraphics[width=\columnwidth]{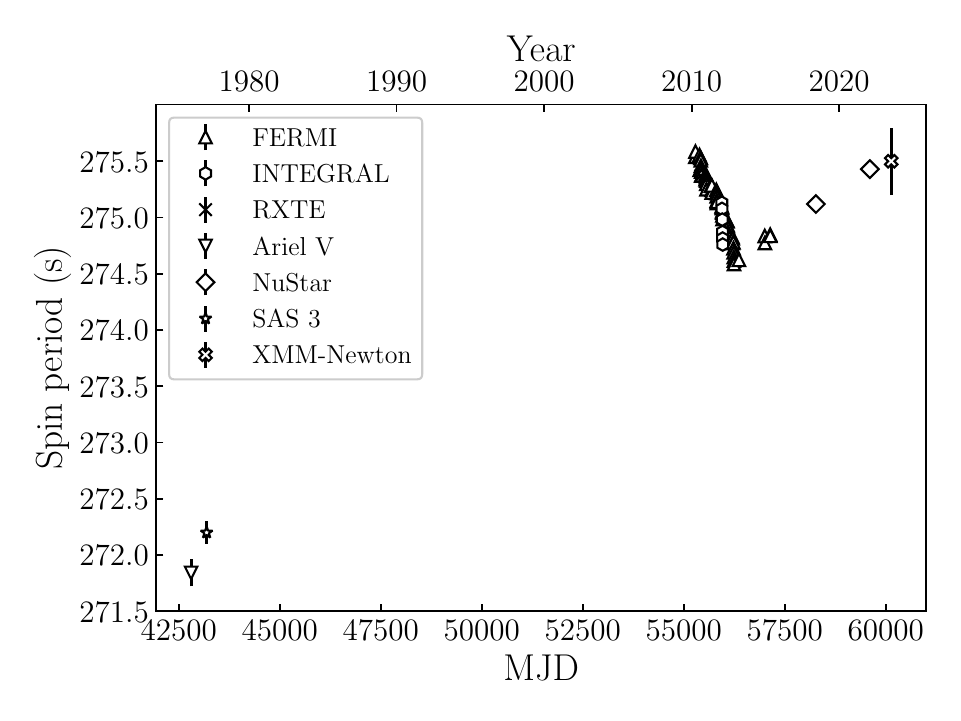}
  \caption{Long-term spin history of GX 304-1 from 1977 February until 2023 July. The reversal in trend from long-term spin-up to long-term spin-down of the X-ray pulsar is discernible between the period 2010-2023.}
 \label{f1}
\end{figure}

We use all the spin period measurements reported in the literature, those measured by the \textit{FERMI}/GBM observations and the spin periods estimated in this work from \textit{NuStar} and \textit{XMM-Newton} observations to construct the long-term spin history of GX 304-1, which is shown in Figure {\ref{f1}}. The spin evolution of this pulsar spans nearly five decades (February 1977 until July 2023).
272 s pulsations from the source were detected by \citet{mcclintock1977discovery} using \textit{SAS-3} observations, which were also detected by \citet{huckle1977discovery} using \textit{Ariel V} observations. Thereafter, the source became quiescent for almost three decades when it was detected using \textit{INTEGRAL} observations \citep{manousakis2008integral}. Pulsations having a periodicity of about 275.5 s was detected by \textit{FERMI}/GBM during the onset of an outburst in 2010 April which suggests that the pulsar spun down by about 3.3 s during the long dormant period. The estimated spin-down rate during this quiescent period is $\sim -4.3 \times 10^{-14}$\,Hz\,s$^{-1}$ which is similar to those detected in other BeXRB pulsars during long ($\sim$yr) dormant periods \citep{malacaria2020ups,chandra2023astrosat}. During the active period since 2010, the pulsar underwent a series of regular outbursts until 2013 before the outburst decayed. The pulsar exhibited spin-up during outbursts at a rate of
$\sim 1 \times 10^{-12}$\,Hz\,s$^{-1}$ \citep{sugizaki2017correlation,malacaria2020ups} and spin-down between outbursts at a rate of $\sim -5 \times 10^{-14}$\,Hz\,s$^{-1}$ \cite{malacaria2020ups}. The long-term spin-up rate during this active episode of the pulsar was estimated to be $\sim 1.3 \times 10^{-13}$\,Hz\,s$^{-1}$ \citep{malacaria2020ups}.

The pulsar was detected by \textit{FERMI}/GBM around MJD 57005 (2014 December 14) and MJD 57008 (2014 December 17) having spin period of about 274.8 s suggesting that the pulsar had undergone a change from a long-term spin-up to a spin-down trend during the preceding quiescent episode of the pulsar. Interestingly, the pulsar was again detected after one orbital period (about 132 d) around MJD 57140 (2015 April 28) and MJD 57143 (2015 May 1) having a similar spin period of about 274.8 s which confirms the onset of spin-down episode of the pulsar. The spin period measured using \textit{NuStar} observation during the low X-ray luminosity state of the pulsar around MJD 58272 (2018 June 3) was 275.12$\pm$0.02 s \citep{escorial2018discovery}. Spin periods estimated in this work using the \textit{NuStar} and the \textit{XMM-Newton} observations from 2022 January and 2023 July corroborates the detection of a long-term spin-down manifestation in this pulsar. The estimated long-term spin-down rate is $\sim -3.4 \times 10^{-14}$\,Hz\,s$^{-1}$ which is similar to those detected during spin-down periods between X-ray outbursts in this source. This suggests that the underlying mechanism of spin-down between regular outbursts observed in this source during  2010-2013 and that during the ongoing quiescent period since 2015 might be the same. Spin variations on long time-scales have been detected in several accretion-powered pulsars  \citep{makishima1988spin,nagase1989accretion, bildsten1997observations,chakrabarty1997correlation,fritz2006torque,camero2009new,inam2009recent,gonzalez2012spin,chandra2021detection}.

\subsection{X-ray activity during change in long-term spin trend}

\begin{figure*}
        \begin{subfigure}[b]{0.48\textwidth}
                \includegraphics[width=\linewidth,keepaspectratio=true]
                {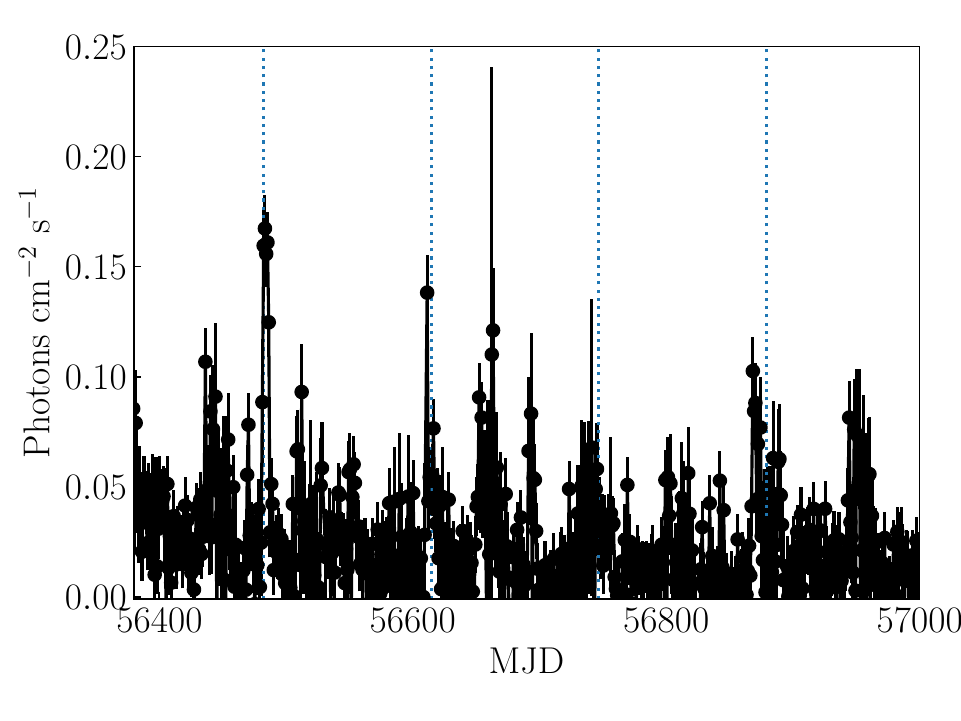}
                \caption{}
        \end{subfigure}
        \begin{subfigure}[b]{0.48\textwidth}
                \includegraphics[width=\linewidth]{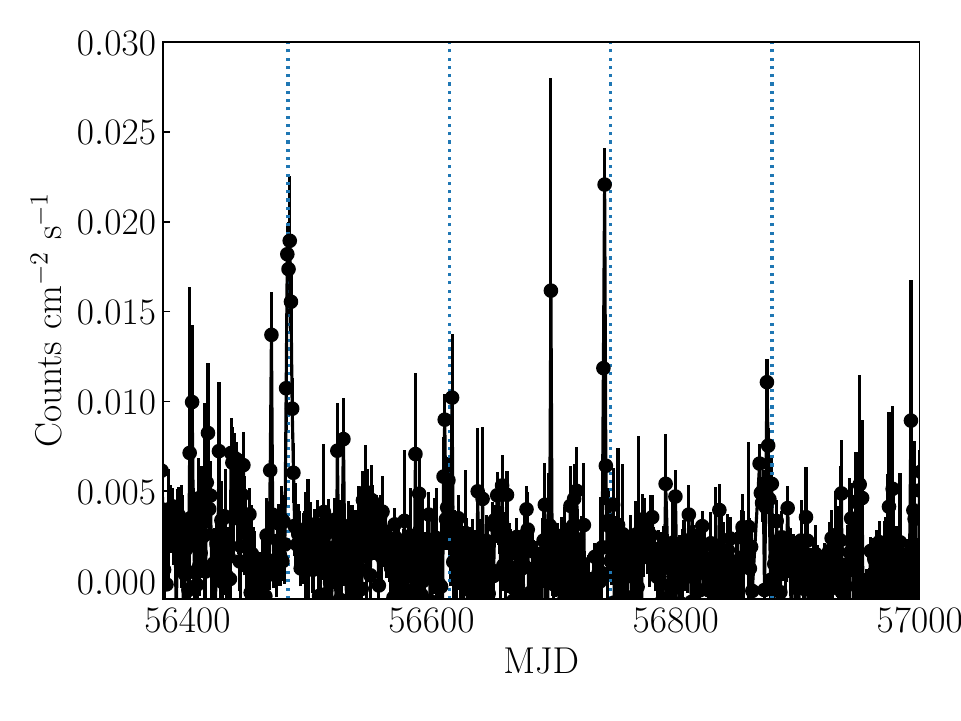}
                \caption{}
        \end{subfigure}
        \caption{(i) \textit{MAXI} one day averaged light curve of GX 304-1 in
the 2-20 keV energy band spanning the duration when the change in long-term spin trend occurred in this pulsar. The dotted vertical lines indicate the epochs of periastron passages of the neutron star. (ii) \textit{Swift}/BAT one day averaged light curve of GX 304-1 in
the 15-50 keV energy band shown during the same duration with times of periastron passages marked with vertical dotted lines.}
        \label{f2}
\end{figure*}

The X-ray light curves in the 2-20 keV and 15-50 keV energy bands from the \textit{MAXI} and \textit{Swift}/BAT monitoring observations of the source are shown in Figure \ref{f2}. The light curves are plotted for the period spanning MJD 56380-57000 (2013 March 29-2014 December 9) when the source underwent spin evolution from a long-term spin-up to spin-down trend and was undetected by \textit{FERMI}/GBM observations. The dotted vertical lines in the figures show the epochs of periastron passages using the orbital
parameters $P_{\rm{orb}}$=132.189 d and $T_0$=MJD 55425.6 \citep{sugizaki2015luminosity}. There is an indication of a weak X-ray brightening of the source during periastron passages around MJD 56482 (2013 July 9), MJD 56614 (2013 November 18), MJD 56746 (2014 March 30), and MJD 56879 (2014 August 10) which is seen in both \textit{MAXI} and \textit{Swift}/BAT observations. The estimated X-ray luminosities during these periastron passages are about ${2}\times 10^{35}$\,erg\,s$^{-1}$, ${7}\times 10^{34}$\,erg\,s$^{-1}$, ${6}\times 10^{34}$\,erg\,s$^{-1}$ and ${2}\times 10^{34}$\,erg\,s$^{-1}$ which suggests that accretion is not quenched during this period. The luminosities are inferred using the luminosities given in \cite{tsygankov2019dramatic} when the source was in a low luminosity state and then scaling the luminosity using the \textit{MAXI} count rate for a given epoch (assuming that there is no spectral evolution at low luminosities which is confirmed by \cite{escorial2018discovery}).

\subsection{Long term photometric observations of the companion star}
\begin{figure}
\centering
  \includegraphics[width=\columnwidth]{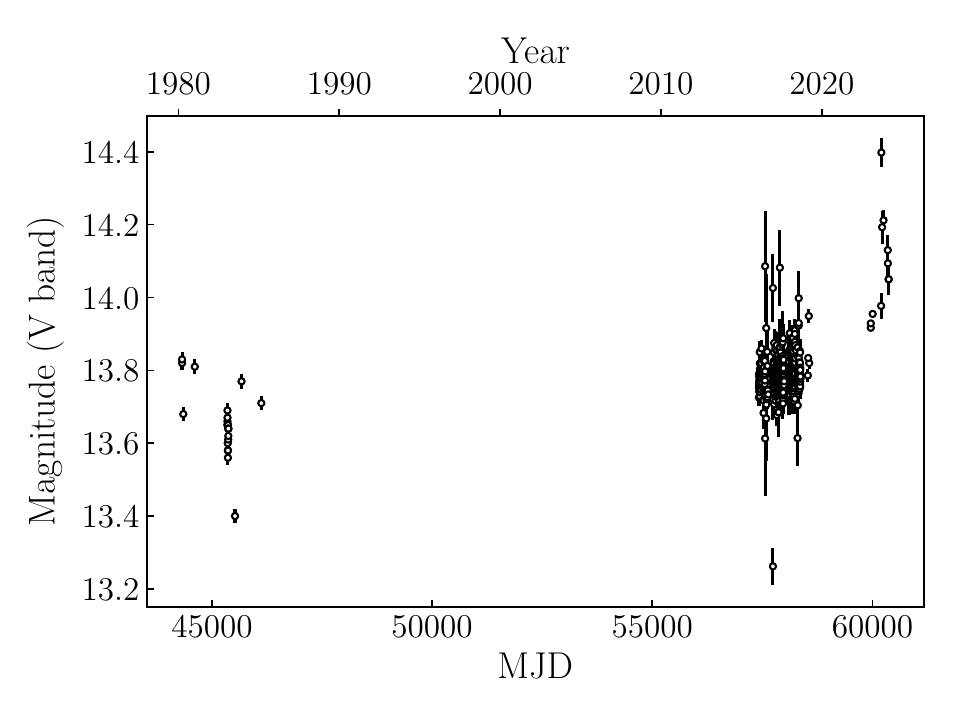}
  \caption{Long-term photometric observations of the companion star of GX 304-1 using observations reported in the literature \citep{corbet1986long,haefner1988optical} spanning the duration MJD 44285.5 until MJD 46121.5 and those obtained from the ASAS-SN optical observations (\url{https://asas-sn.osu.edu/}), AAVSO (\url{https://www.aavso.org/}) and \textit{Swift}/UVOT.}
 \label{f3}
\end{figure}

The long-term photometric observations of the companion star V850 Cen is shown in Figure \ref{f3}.
We have plotted the V-band magnitudes reported in the literature \citep{corbet1986long,haefner1988optical} during the period MJD 44285.5 (1980 February 16) until MJD 46121.5 (1985 February 25) along with those obtained from the ASAS-SN optical observations (\url{https://asas-sn.osu.edu/}), AAVSO (\url{https://www.aavso.org/}) and \textit{Swift}/UVOT. The averaged V-band magnitude during the period spanning almost five years from 1980-85 was $\sim$13.6, while that during the period MJD 57423.7 (2016 February 5) to MJD 60367.8 (2024 February 27)
was $\sim$13.9. Optical brightening of the companion star in BeXRBs has been associated with X-ray outbursts in these systems \citep{corbet1985optical,negueruela1997multiwavelength,reig2007x,caballero2016activity,coe20242022}.

The increase in averaged V-band magnitude by about 0.3 suggests that the companion star has become relatively less active recently due to which there have likely been no episodes of mass ejection from the companion star that feeds the relatively bright X-ray outbursts in this binary pulsar. This is confirmed by the weak X-ray detection of the pulsar on MJD 57411 (2016 January 24) and MJD 57525 (2016 May 17) at a flux level of 23$\pm$4 mCrab and 22$\pm$14 mCrab respectively using \textit{MAXI} observations \citep{nakajima2016maxia,nakajima2016maxib} after which the pulsar has remained dormant. Unfortunately, no regular optical monitoring observations of the source is available during the period when the pulsar switched from a long-term spin-up trend to spin-down except for one optical spectroscopic observation (around MJD 56763 (2014 April 16)) using the the 3.9 m \textit{Anglo Australian Telescope} (\textit{AAT}) which suggested that the radius of the decretion disc around the companion star had shrunk significantly (by almost a factor of 2) compared to the epochs when the pulsar was moderately active in X-rays \citep{malacaria2017optical}.

\subsection{Long term ultraviolet observations of the companion star}
\begin{figure}
\centering
  \includegraphics[width=\columnwidth]{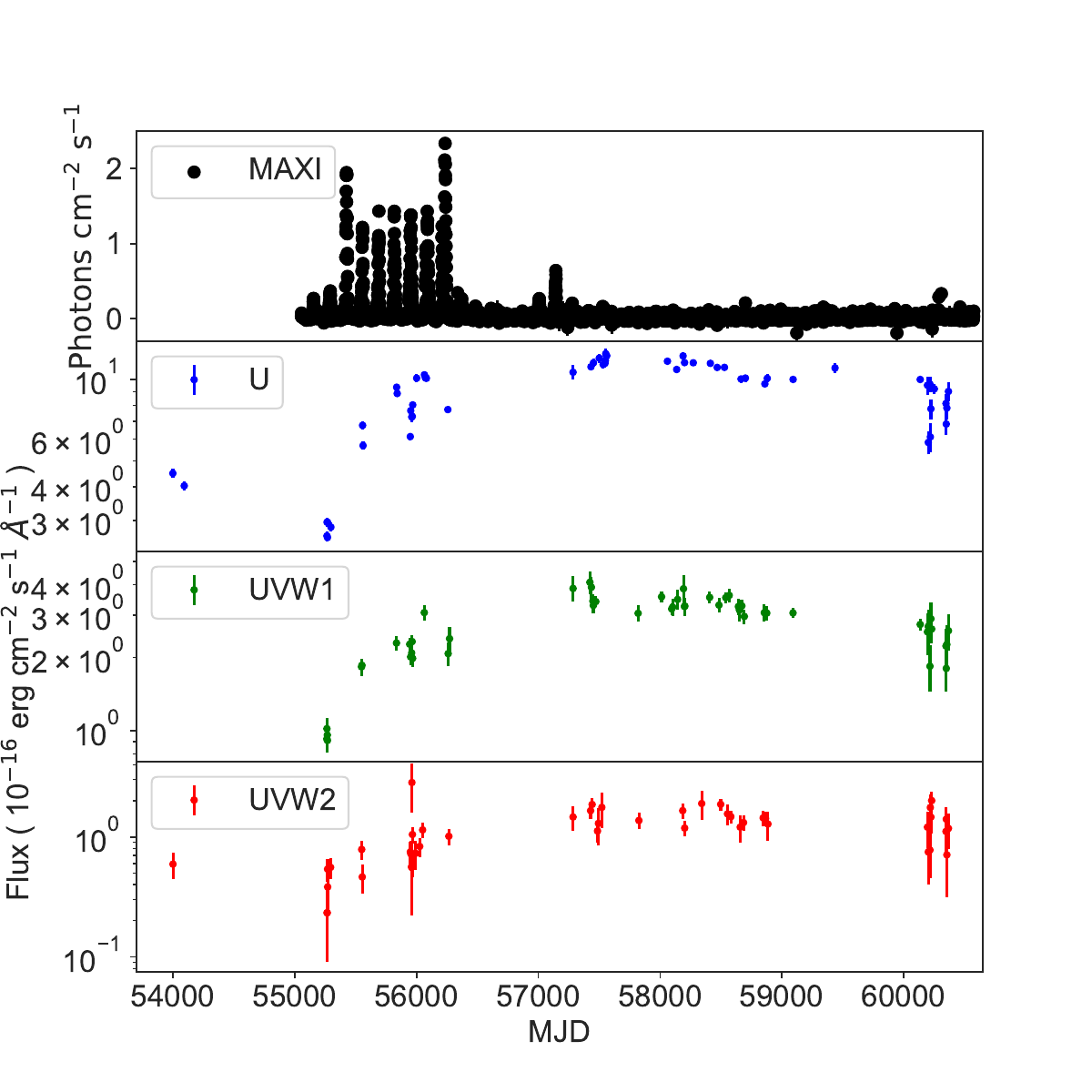}
  \caption{Long-term X-ray and ultraviolet observations of GX 304-1 as observed with \textit{MAXI} and \textit{Swift}/UVOT respectively. We show, from top to bottom, the one-day averaged X-ray count rate in the 2-20 keV band from \textit{MAXI} and the UV fluxes in the U, UVW1 and UVW2 bands from \textit{Swift}/UVOT.}
 \label{f30}
\end{figure}

The X-ray (2-20 keV) and ultraviolet light curves for GX 304-1 obtained from \textit{MAXI} and \textit{Swift}/UVOT observations are shown in Figure \ref{f30}. The \textit{Swift}/UVOT light curves show the inferred fluxes in the U, UVW1 and the UVW2 filters. The ultraviolet fluxes of the source show a marked gradual increase around MJD 55300 (2010 April) until around MJD 56000 (2012 March) in all three filters (U, UVW1 and UVW2) which interestingly coincides with the re-kindling of the X-ray outbursts detected in this pulsar as observed from simultaneous \textit{MAXI} X-ray light curve. The increase in the ultraviolet fluxes in the U, UVW1 and the UVW2 filters are about a factor of 2.3, 2 and 1.5 during the dormant period (around MJD 55268) compared to that during the beginning of X-ray outbursts (around MJD 55550). It also seems that the increase in the ultraviolet activity of the companion star preceded the onset of bright X-ray outbursts by about half a year. A similar increase in the ultraviolet flux in the UVW1 band preceded the onset of X-ray outbursts in the BeXRB Swift J004516.6-734703 by about a year \citep{kennea2020newly}. The increase in UV flux was attributed to the formation of a circumstellar disc around the companion star in Swift J004516.6-734703 \citep{kennea2020newly}. It is likely that the increase in the UV signatures in GX 304-1 might also indicate the resurgence of the formation of a decretion disc around the companion star which was big enough to trigger bright X-ray outbursts in the binary after about half a year. After the regular X-ray outbursts ceased in GX 304-1, the ultraviolet fluxes in the three bands have shown little long-term variation with no indication of returning to the pre-outburst level. However, there is suggestion of a decrease in the ultraviolet fluxes around MJD 60300 (2023 December) on short timescales.

\subsection{Long-term quiescent X-ray activity of GX 304-1 using \textit{Swift}/XRT}

\begin{figure*}
        \begin{subfigure}[b]{0.48\textwidth}
                \includegraphics[width=\linewidth,keepaspectratio=true]
                {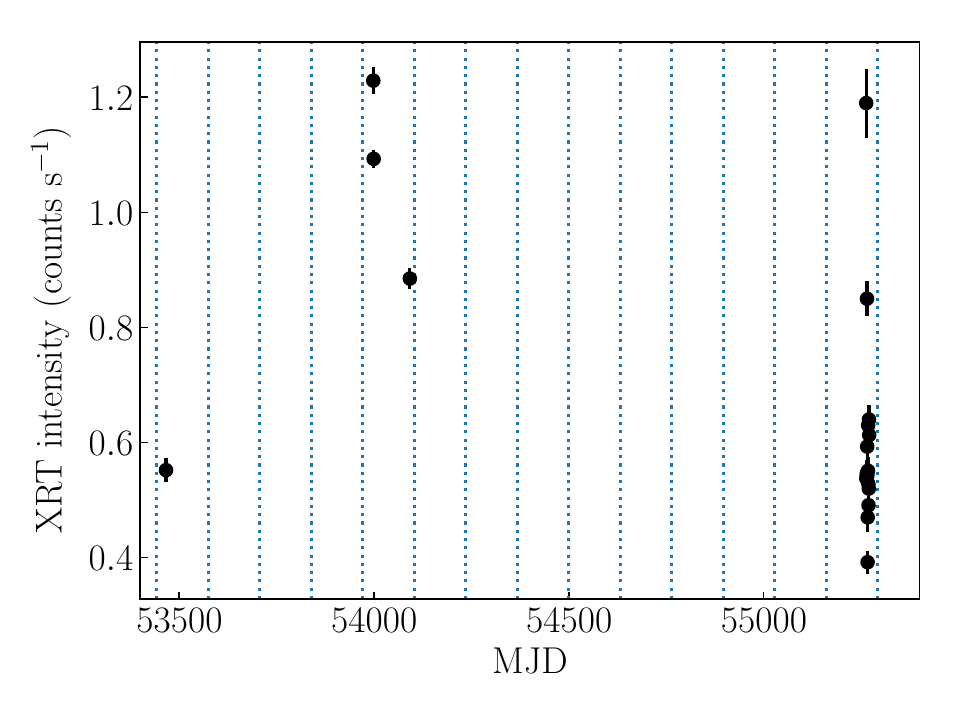}
                \caption{}
        \end{subfigure}
        \begin{subfigure}[b]{0.48\textwidth}
                \includegraphics[width=\linewidth]{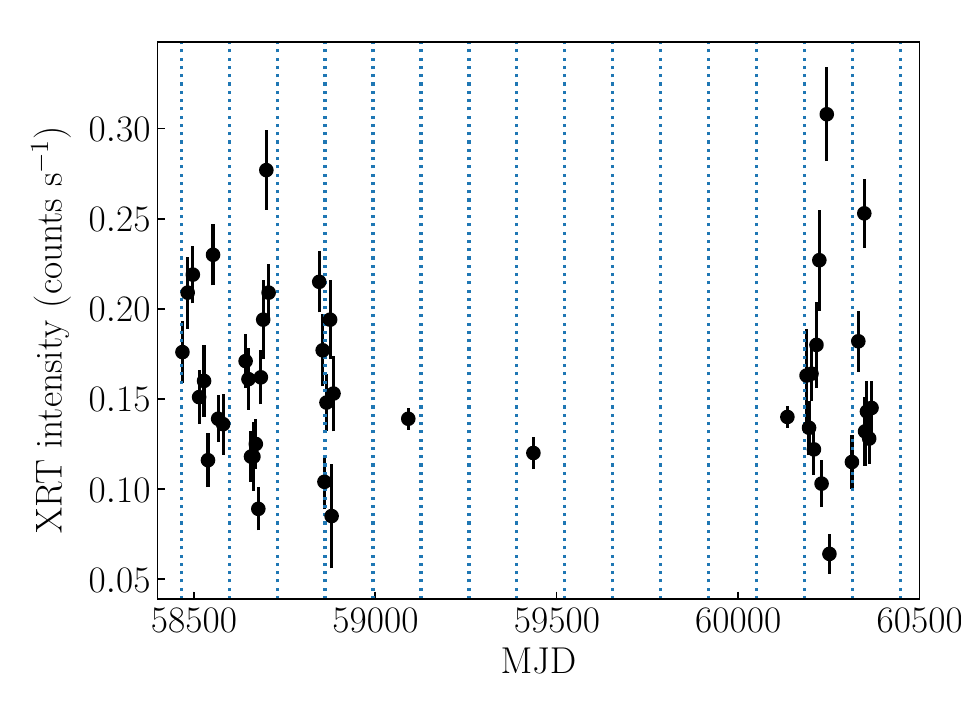}
                \caption{}
        \end{subfigure}
        \caption{ (i) \textit{Swift}/XRT light curve (0.3-10 keV energy band) of GX 304-1 spanning the quiescent phase for the duration 2005 April 6 (MJD 53466) until 2010 March 15 (MJD 55270.86). The vertical dotted lines indicate the epochs of periastron passages of the neutron star. The averaged XRT count rate during this duration is about 0.7 $\rm{counts ~s^{-1}}$. (ii) Same as (i) for the duration spanning nearly five years from 2018 December 17 (MJD 58469) until 2024 February 27 (MJD 60367.8). The averaged XRT count rate during this duration is about 0.2 $\rm{counts ~s^{-1}}$.}
        \label{f0}
\end{figure*}

The long-term \textit{Swift}/XRT light curve of GX 304-1 in the 0.3-10 keV band is shown in Figure {\ref{f0}}(i) for the duration spanning
2005 April 6 (MJD 53466) until 2010 March 15 (MJD 55270.86) and in Figure {\ref{f0}}(ii) for the period from 2018 December 17 (MJD 58469) until 2024 February 27 (MJD 60367.8). The count rates have been averaged for a given epoch. It is observed that the XRT count rate during the period between MJD 53466-55270.86
varied from about $\sim$0.4$~\rm{counts ~s^{-1}}$ to $\sim$1.2 $\rm{counts ~s^{-1}}$ showing a dynamic range of about 3. The averaged count rate during this period was about 0.7 $\rm{counts ~s^{-1}}$. This suggests that the pulsar was moderately active in X-rays but did not undergo an outburst as the peak XRT count rates even during weak outbursts in 2016 February was about 2-4 $\rm{counts ~s^{-1}}$ \citep{escorial2018discovery}.

The XRT count rate spanning the duration 2018 December 17 until 2024 February 27 varied from about 0.06-0.3 $\rm{counts ~s^{-1}}$ (having a dynamic range of about 5) which suggests that the pulsar was in a low X-ray activity state. This quiescent behaviour of the pulsar is similar to that observed from 2016 June until 2018 October when the XRT count rate varied in the range of about 0.1-0.25 $\rm{counts ~s^{-1}}$ \citep{escorial2018discovery}. The averaged count rate during this period was about 0.2 $\rm{counts ~s^{-1}}$. As seen in Figure {\ref{f0}}(ii), the pulsar also did not exhibit any clear enhancement in X-ray activity at periastron passages. This intriguing low luminosity state of the source has now lasted for about seven years since September 2017. Similar long-term quiescent episodes on nearly decadal timescales has been detected in other Be/X-ray binaries such as 1A 0535+262
,  GS 0834-430, RX J0209.6-7427, XTE J1946+274, 2S 1417-624, KS 1947+300, and RX J0440.9+4431   \citep{usui2012outburst,miyasaka2013nustar,gupta2019nustar,chandra2020study,chandra2023astrosat,liu2023multiwavelength}.

\section{Discussions}
In the following section, we discuss plausible mechanisms that can lead to a long-term spin-down and low luminosity state in this  pulsar.

\subsection{Quasi-spherical accretion in GX 304-1}

\begin{figure}
\centering
  \includegraphics[width=\linewidth]{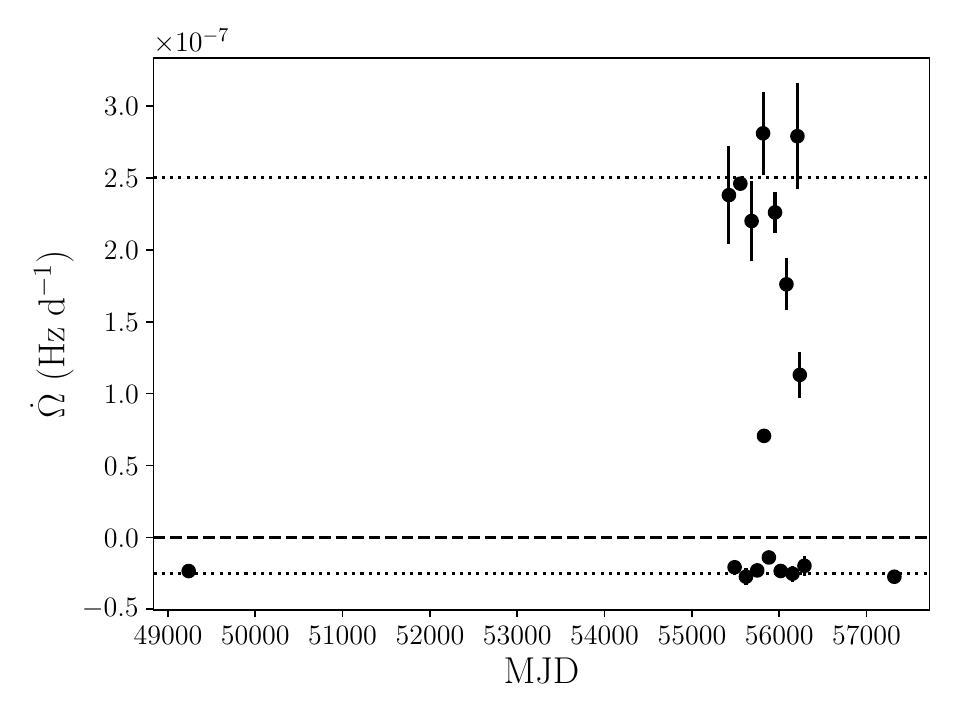}
  \caption{Plot showing estimated spin-up and spin-down rates spanning the duration of about five decades. The estimated spin-up and spin-down rates using the quasi-spherical settling accretion theory are shown using horizontal dotted lines \citep{shakura2012theory,shakura2014theory,postnov2015spin}.}
 \label{f9}
\end{figure}

The long-term spin-down in this pulsar can be explained using the theory of quasi-spherical accretion from the stellar wind of the companion star \citep{shakura2012theory,shakura2014theory}. The conditions for the applicability of this model such as slow spin period and X-ray luminosity $< {4}\times 10^{36}$\,erg\,s$^{-1}$ are fulfilled by GX 304-1 observations. The spin period of this pulsar is about 275 s and the inferred X-ray luminosity during the period when the spin evolution changed from a long-term spin-up to the spin-down regime is about ${2}\times 10^{35}$\,erg\,s$^{-1}$ except during bright outbursts of the source when the luminosity reaches about $10^{37}$\,erg\,s$^{-1}$ \citep{yamamoto2011discovery,klochkov2012outburst,malacaria2015luminosity,sugizaki2015luminosity,jaisawal2016suzaku}. The inferred luminosity during the low state of the pulsar from \textit{XMM-Newton} observations (when the pulsar was spinning down) is about $2.5 \times 10^{33}$\,erg\,s$^{-1}$ in the 0.5-10 keV energy band. The spin-up during outbursts and spin-down on short time-scales in between regular outbursts have been explained using this model by \cite{postnov2015spin}. Using the quasi-settling accretion theory, the estimated spin-down rate is given by \citep{postnov2015spin},

\begin{equation}
 {\dot{\omega}^*}_{sd}\sim 10^{-8} \frac{Hz}{d} \Pi_{sd} \mu^{13/11}_{30}\dot{M}^{3/11}_{16}\left({\frac{P^*}{100 ~\rm{s}}}\right)^{-1},
\end{equation}

where $\mu_{30}=\mu/10^{30} [\rm{G~cm^3}]$ is the dipole magnetic moment given by $\mu=\rm{BR^3/2}$ where R is the radius of the neutron star having a typical value of 10 km, $\dot{M}_{16}=\dot{M}/10^{16}[\rm{g~s^{-1}}]$ is the mass accretion rate onto the neutron star and $P^*$ is the equilibrium spin period of the neutron star. We obtain $\dot{{\omega}^*}_{sd} \sim -3.1 \times 10^{-8}$ Hz/d (using $\Pi_{sd} \sim 4.6$ \citep{shakura2012theory, shakura2014theory, postnov2015spin}, $\dot{M}_{16}=0.22$ using $\mathrm{L_X}=0.1\dot{M}\mathrm{c^2}$ and $\mathrm{L_X}={2}\times 10^{35}$\,erg\,s$^{-1}$, $\mu_{30}=2.35$ and $P^{*}_{eq}=275 ~$s) which is slightly larger than the estimated
long-term spin-down rate by a factor of $\sim$1.7. Figure \ref{f9} shows the observed spin-up/down rates obtained from a linear fit of spin evolution of the pulsar during spin-up/down episodes.

The spin-up rate estimated during outbursts using the quasi-settling accretion theory is $\sim 2.5 \times 10^{-7}$ Hz/d \citep{postnov2015spin}. The dotted horizontal lines in Figure \ref{f9} show the estimated spin-up/spin-down rates using the quasi-spherical settling accretion theory. The long-term spin-down rates are remarkably similar to the short time-scale spin-down rates estimated between outbursts in this source which suggests that both the short-term and long-term spin-down might be caused by the same mechanism in this pulsar. The long-term spin-up rate is about a factor of 4 smaller than the spin-up rates estimated during X-ray outbursts detected in this pulsar. Thus, this model can estimate the spin-down rate within a factor of $\sim$1.7 and may qualitatively explain the long-term spin-down in this pulsar. Quasi-spherical accretion has been used to explain the long-term spin-down episodes in GX 1+4 and Vela X-1 \citep{shakura2012theory,chandra2021detection}.

\subsection{Tug of war between spin-up and spin-down torques}

It is observed around MJD 56000 (ref. Fig. \ref{f9}) that the spin-up rates estimated during outbursts seem to diminish systematically with time (except for one instance of rapid spin-up during a giant outburst detected in this pulsar) by a factor of about 2 as the pulsar changes trend from a long-term spin-up to spin-down. However, the spin-down rates inferred in between the outbursts remain nearly the same during this period (Fig. \ref{f9}). The estimated peak X-ray flux in the 2-10 keV energy band during the spin-down episodes in between the outbursts diminished by a factor of about 1000 compared to the flux during the spin-up regimes (ref. Table 1 in \cite{postnov2015spin}) which suggests that the accretion rate onto the neutron star ($\dot{M}$) decreased by the same factor during spin-down phases. The spin-up torque ($K_{\rm su}$) acting on the neutron star is given by \citep{pringle1972accretion},

\begin{equation}
 K_{\rm su} = \dot{M} (GM_{\rm ns} r_{\rm m})^{1/2},
\end{equation}

where G is the gravitational constant, $M_{\rm ns}$ is the mass of the neutron star and $r_{\rm m}$ is the radius at which the effective pressure in the accretion disc equals the magnetic pressure. As the accretion rate drops during the spin-down regimes, consequently the spin-up torque also decreases significantly during this period. The spin evolution of an accretion-powered pulsar is given by,

\begin{equation}
 2 \pi I \dot{\nu} = K_{\rm su} - K_{\rm sd},
\end{equation}

where $I$ is the moment of inertia of the neutron star and $K_{\rm sd}$ is the spin-down torque acting on the neutron star. Assuming that the spin-down torque acting on the neutron star is nearly constant (which is observed in Fig. \ref{f9} that the estimated spin-down torques in between outbursts are almost the same), the spin-down torque would eventually overtake the decreasing spin-up torque (after around MJD 56200) leading to a long-term spin-down in the pulsar.

The estimated averaged X-ray luminosities (in the 0.5-100 keV energy range using the method described earlier) during the period MJD 57005-57008 and MJD 57140-57143 when the pulsar was already spinning down was $\sim {4.4}\times 10^{35}$\,erg\,s$^{-1}$ and $\sim {9.6}\times 10^{35}$\,erg\,s$^{-1}$ respectively. These luminosities are about a factor of 10-100 smaller than those detected during bright X-ray outbursts in this pulsar which suggests that the increase in accretion rate during this period did not lead to a sufficient increase in the spin-up torque to change the long-term spin-down trend of the pulsar. The estimated luminosities after about a year around MJD 57423 and MJD 57506 were about ${1.2}\times 10^{35}$\,erg\,s$^{-1}$ and ${2.8}\times 10^{34}$\,erg\,s$^{-1}$ respectively \citep{rouco2016swift}. From long-term X-ray monitoring observations of the source spanning the duration around MJD 58000-58420, the estimated luminosities were found to lie in the range  $\sim {0.8-1.9}\times 10^{34}$\,erg\,s$^{-1}$ \citep{rouco2016swift}. Recent \textit{Swift}/XRT, \textit{XMM-Newton} and \textit{NuStar} observations analysed in this work confirm the prolonged low state of the pulsar. The relatively low luminosity of the pulsar since the last bright outburst detected in 2013 suggests relatively weak spin-up torque acting on the pulsar and the continued dominance of the spin-down torque over the spin-up torque. The estimated spin period of the source (275.12$\pm$0.02 s) around MJD 58272 \citep{rouco2016swift} and those obtained from \textit{NuStar} and the \textit{XMM-Newton} observations in this work confirms this proposition as the pulsar continues to spin-down on long time-scales.

The long-term spin-down of the pulsar ($\sim -4.3 \times 10^{-14}$\,Hz\,s$^{-1}$) changed to a short-term spin-up trend after the onset of outbursts in 2010 ($\sim 1.3 \times 10^{-13}$\,Hz\,s$^{-1}$) with the averaged spin-up rate being an order of magnitude higher than the long-term spin-down rate. This suggests that the average momentum imparted by the
accreted material during the spin-up phase was greater than the momentum lost due to magnetic breaking over the long-term spin-down phase lasting about 28 years which is also observed in SMC X-3 \citep{townsend20172016}. We now discuss possible mechanisms to explain the enigmatic long-term low luminosity state of the pulsar since 2018 December.

\subsection{Propeller effect}
\begin{figure*}
        \begin{subfigure}[b]{0.48\textwidth}
                \includegraphics[width=\linewidth,keepaspectratio=true]
                {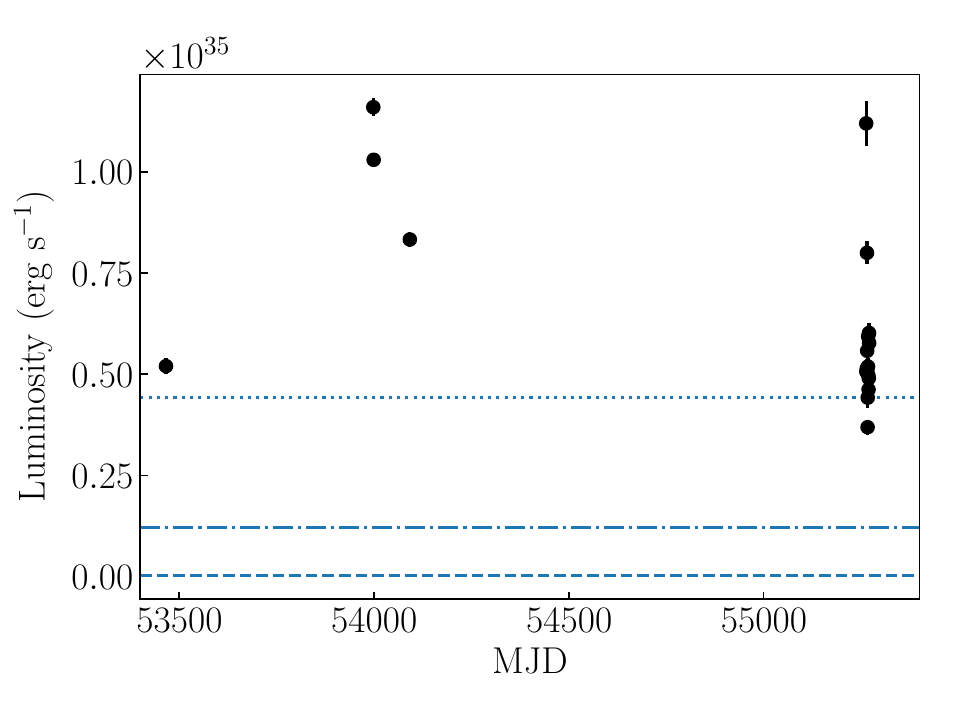}
                \caption{}
        \end{subfigure}
        \begin{subfigure}[b]{0.48\textwidth}
                \includegraphics[width=\linewidth]{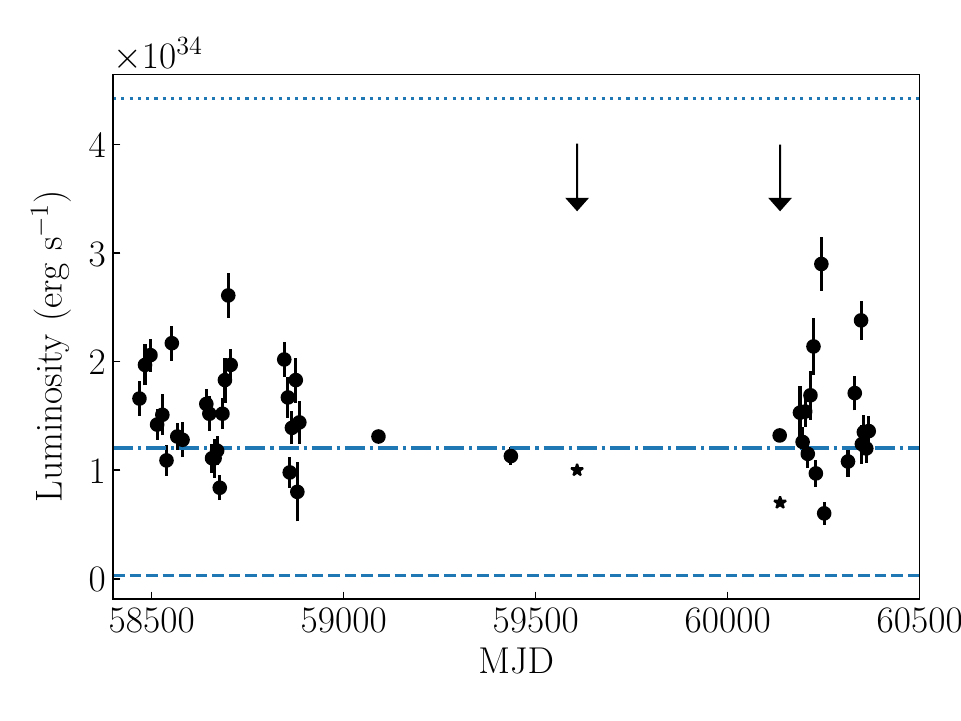}
                \caption{}
        \end{subfigure}
        \caption{ (i) Plot showing estimated X-ray luminosity (0.5-100 keV) of GX 304-1 spanning the duration 2005 April 6 (MJD 53466) until 2010 March 15 (MJD 55270.9). The horizontal dashed line shows the estimated limiting luminosity for the propeller effect to set in. The dotted and dash-dotted horizontal lines show the limiting luminosity for accretion to occur from a cold disc. The averaged X-ray luminosity (0.5-100 keV) during this duration is about $6.3 \times 10^{34}$\,erg\,s$^{-1}$. (ii) Same as (i) for the duration spanning from 2018 December 17 (MJD 58469) until 2024 February 27 (MJD 60367.8). The averaged X-ray luminosity (0.5-100 keV) during this duration is about $1.5 \times 10^{34}$\,erg\,s$^{-1}$. The downward arrows show the epoch of the \textit{NuStar} and the \textit{XMM-Newton} observations used in this study and the estimated 0.5-100 keV unabsorbed luminosities for these epochs are shown by stars.}
        \label{f10}
\end{figure*}

  The XRT count rates shown in Figure {\ref{f0}} are converted into the corresponding 0.5-100 keV luminosities (Fig. {\ref{f10}}) using the spectral study and luminosities from \citet{tsygankov2019dramatic} (when the source was faint having an XRT count rate of about 0.13 $\rm{counts ~s^{-1}}$) and then  scaling our observed XRT count rates which are listed in Tables {\ref{tab1}} and {\ref{tab2}}. An assumption is made during this conversion that the spectrum of the pulsar does not change during faint state which is confirmed by \citet{escorial2018discovery}. The estimated luminosities shown in Figure {\ref{f10}}(i) for the duration 2005 April 6 (MJD 53466) until 2010 March 15 (MJD 55270.9) varied in the range of $\sim {3.7-11.2}\times 10^{34}$\,erg\,s$^{-1}$ showing variation by a factor of about 3. The luminosities for the duration spanning 2018 December 17 (MJD 58469) until 2024 February 27 (MJD 60367.8)  is shown in Figure {\ref{f10}}(ii) showing luminosities in the range  of $\sim {0.6-2.9}\times 10^{34}$\,erg\,s$^{-1}$, which suggests that the source has been quasi-stable in a low luminosity state for the last five years.  The pulsar was in a slightly higher low luminosity regime during 2005 April 6-2010 March 15. Accreting pulsars may switch to the propeller regime at low mass accretion rates caused by the centrifugal barrier if the velocity of the rotating magnetosphere exceeds the local Keplerian velocity \citep{illarionov1975number}. The critical luminosity for the onset of the propeller effect ($L_{\rm prop}$) is given by \citep{tsygankov2017stable},

\begin{equation}
  L_{\rm prop}
\simeq \frac{GM\dot{M}_{\rm prop}}{R} \simeq 4 \times 10^{37} k^{7/2}
B_{12}^2 P_s^{-7/3} M_{1.4}^{-2/3} R_6^5 \,\textrm{erg s$^{-1}$} , \label{eq4}
\end{equation}

where M, R, $P_s$ and B are the mass, radius, spin period and magnetic field of the neutron star respectively. $\dot{M}_{\rm prop}$ is the mass-accretion rate onto the neutron star.  The factor $k$ is the ratio of the magnetospheric radius and the Alfv\'en radius which in the case of disc accretion is taken to be $k=0.5$ \citep{ghosh1978disk}. Using $M=1.4 ~M_{\odot}$, R=10 km, $P_s \sim 275.4 ~s$, B=$4.7 \times 10^{12} ~G$ \citep{yamamoto2011discovery} and k=0.5 (assuming disc accretion), the estimated $L_{\rm prop}$ is $\sim 3.2\times 10^{32}$\,erg\,s$^{-1}$ which is shown by a dashed horizontal line in Figure {\ref{f10}}(i) and (ii).

The magnetic field (B=$4.7 \times 10^{12} ~G$ \citep{yamamoto2011discovery}) used in estimating the limiting luminosity due to the onset of the propeller effect using equation (\ref{eq4}) was obtained from the detection of cyclotron resonance scattering feature (CRSF) in the spectra. The magnetic field can also be estimated independently by using \citep{christodoulou2016x},

\begin{equation}
B = \left(2\pi^2 \xi^7\right)^{-1/4}
\sqrt{\frac{G M I}{R^6} |\dot{P_S}|} \, ,
\end{equation}

where $\xi$ is a dimensionless parameter which is the ratio of the inner edge of the accretion disc and the magnetospheric radius \citep{ghosh1979accretion,wang1996location,christodoulou2016x}, M and R are the mass and the radius of the neutron star respectively, I is the moment of inertia of the neutron star given by $I=\frac{2MR^2}{5}$ and $\dot{P_S}$ is the rate of spin change of the pulsar during outbursts. Using $\xi$=1, $M=1.4 ~M_{\odot}$, R=10 km, $\dot{P_S} \sim 3.1 \times 10^{-8}$\,s\,s$^{-1}$ \citep{postnov2015spin} and $G=6.67\times 10^{-8}$~cm$^3$~g$^{-1}$~s$^{-2}$, the estimated magnetic field of the neutron star is $\sim 3.8 \times 10^{13} ~G$, which is higher by a factor of about 8 than that inferred from the detection of cyclotron line in the spectra. Using B$\sim 3.8 \times 10^{13} ~G$, the estimated luminosity for the onset of the propeller effect is  $\sim 2.6\times 10^{33}$\,erg\,s$^{-1}$, which is about an order of magnitude higher than that estimated earlier. The magnetic field of the neutron star can also be estimated independently using the models given by \cite{ghosh1979accretion} and \cite{kluzniak2007magnetically} which are applicable to disc-fed systems. The Ghosh and Lamb model \citep{ghosh1979accretion} is applicable to systems irrespective of whether they have achieved spin equilibrium and the model predicts \citep{klus2014spin}

\begin{equation}
  -\dot{P}=5.0\times10^{-5} \mu_{30}^{2/7} n(\omega_{s}) R_{6}^{6/7}\left(\frac{M}{M_\odot}\right)^{-3/7}I_{45}^{-1} (PL_{37}^{3/7})^2,
\end{equation}

where $\dot{P}$ is the long-term spin derivative of the neutron star (in s yr$^{-1}$),  I is the moment of inertia of the neutron star and $n(\omega_{s})$ is the dimensionless accretion torque and
depends on the fastness parameter $\omega_{s}$ \citep{ghosh1979accretion,klus2014spin}. Using $\dot{P}\sim 0.08$ s yr$^{-1}$, $n(\omega_{s})\sim$1, $R=10^6$ cm, $M=1.4 ~M_{\odot}$, $I_{45}$=1.92\,g\,cm$^3$, P=275 s and $L \sim 2\times 10^{37}$\,erg\,s$^{-1}$, the magnetic field is estimated to be $\sim 5.5 \times 10^{6} ~G$, which is lower by a factor of about $10^6$ than that inferred from the detection of cyclotron line in the spectra.  Using the Kluzniak and Rappaport model \citep{kluzniak2007magnetically}

\begin{equation}
-\dot{P}=8.2\times10^{-5}\mu{_{30}}^{2/7} g(\omega_{s}) R_{6}^{6/7}\left(\frac{M}{M_\odot}\right)^{-3/7}I_{45}^{-1} (PL_{37}^{3/7})^2,
\end{equation}

\textbf{where $g(\omega_{s})$ depends on the fastness parameter $\omega_{s}$ and is nearly equal to unity.
Using $\dot{P}\sim 0.08$\,s\,yr$^{-1}$,
$g(\omega_{s})\sim$1,
$R=10^6$ cm, $M=1.4 ~M_{\odot}$,
$I_{45}$=1.92\,g\,cm$^3$, P=275 s and
$L \sim 2\times 10^{37}$\,erg\,s$^{-1}$,
the magnetic field estimated using this model is $\sim 9.8\times 10^{5} ~G$, which is similar to that estimated using
the Kluzniak and Rappaport model by within a factor of about 6. Similar lower magnetic field estimates compared to that inferred from cyclotron lines were obtained for accreting pulsars (not near spin equilibrium) by \cite{klus2014spin} using the Ghosh and Lamb model \citep{ghosh1979accretion} as well as the Kluzniak and Rappaport model \citep{kluzniak2007magnetically}. The magnetic field estimates obtained using the Ghosh and Lamb model and the Kluzniak and Rappaport model would reduce the threshold luminosity for the onset of the propeller effect by a factor of about $10^6$ compared to that obtained using the magnetic field estimated using the cyclotron line.}

\textbf{The estimated luminosities of the pulsar during the faint state (Fig. \ref{f10}) are well above the threshold luminosity (at least by a factor of about 19 for $L_{\rm{prop}} \sim 3.2\times 10^{32}$\,erg\,s$^{-1}$ and by a factor of about 3 for $L_{\rm{prop}} \sim 2.6\times 10^{33}$\,erg\,s$^{-1}$)} for the propeller effect to set in. Another telltale manifestation of the onset of the propeller effect is the sudden decrease in the luminosity of the pulsar as observed in 4U 0115+63 and V0332+53 \citep{campana2001transient,tsygankov2016propeller} which is not observed in GX 304-1. In addition, the detection of pulsations from the source using \textit{NuStar} and \textit{XMM-Newton} observations indicates that accretion is still continuing at low luminosities. Hence, the long-term low luminosity regime of the pulsar cannot be explained using the propeller effect.

\subsection{Sustained accretion from a cold disc?}

The observed long-term low luminosity regime of the pulsar may be explained using accretion from a \textquotedblleft cold disc\textquotedblright  ~wherein for low accretion rates well above the propeller regime, the temperature of the disc may fall below the Hydrogen ionization temperature of about 6500 K \citep{tsygankov2017stable}. In this case, the matter in the disc is non-ionized (referred to as \textquotedblleft cold disc\textquotedblright) and accretion can proceed through the cold disc \citep{tsygankov2017stable}. Two criteria need to be satisfied for accretion from the cold disc. Firstly, the accretion rate has to be sufficiently high to overcome the centrifugal barrier which implies that the luminosity of the source should be higher than the threshold luminosity for the propeller effect to take over which is satisfied during the \textit{Swift}/XRT, \textit{NuStar} and \textit{XMM-Newton} observations of GX 304-1 (ref. Fig. {\ref{f10}}). Secondly, the luminosity of the source should be lower than the luminosity for stable accretion from a cold disc ($L_{\rm cold}$) given by \citep{tsygankov2017stable},

\begin{equation}
   L_{\rm cold}  =  9\times 10^{33}\,k^{1.5}\,M_{1.4}^{0.28}\,R_6^{1.57}\,B_{12}^{0.86}~~~{\rm erg\,s^{-1}}.\label{eq5}
\end{equation}

The above threshold for $L_{\rm cold}$ assumes that the disc temperature is highest at the magnetospheric radius and as a result the disc temperature for any radius lesser than the magnetospheric radius is less than 6500 K \citep{tsygankov2017stable}. Using $M=1.4 ~M_{\odot}$, R=10 km, B=$4.7 \times 10^{12} ~G$ \citep{yamamoto2011discovery} and k=0.5 (assuming disc accretion), the estimated $L_{\rm cold}$ is $\sim 1.2\times 10^{34}$\,erg\,s$^{-1}$ which is shown by a dash-dotted horizontal line in Figure {\ref{f10}}(i) and (ii).

The observed 0.5-100 keV luminosities of the source lies within a factor of about 2-3 times the $L_{\rm cold}$ estimated for the source during the period since December 2018 and within a factor of about 3-10 during the period spanning 2005 April 6 (MJD 53466) until 2010 March 15 (MJD
55270.9). It should be noted that the equation (\ref{eq5}) does not take into account the interaction between the pulsar magnetosphere and the accretion disc and a more accurate threshold for $L_{\rm cold,acc}$ is given by \citep{tsygankov2017stable},

\begin{equation}
     L_{\rm cold,acc} \simeq 7\times 10^{33}\!\!A^{-7/13}\,k^{21/13}\,M_{1.4}^{3/13}\,R_6^{23/13} 
      B_{12}^{12/13}\,T_{6500}^{28/13}\ {\rm erg\,s^{-1}},
\end{equation}

where A is given by,

\begin{equation}
 A  =
     \left\{ \begin{array}{ll}
       \strut\displaystyle
       0.057\,\beta^{-6}, & {\rm if}~~~\beta \ge \frac{\sqrt{3}}{2},\\
       1-\beta, & {\rm if}~~~\beta < \frac{\sqrt{3}}{2},
       \end{array} \right.
\end{equation}

and $T_{6500}=T_{\rm eff}/6500\,{\rm K}$. The parameter $\beta$ indicates the location in the accretion disc where the viscous stress impacting the temperature distribution of the disc disappears \citep{tsygankov2017stable}. For $\beta$ of unity, the viscous stress disappears at the magnetospheric radius while for $\beta$=0, the stress disappears at a radius much smaller than the magnetospheric radius \citep{tsygankov2017stable}. Assuming that the viscous stress disappears at the magnetospheric radius (i.e. $\beta$=1) and using $T_{\rm eff}$=6500 K, the  threshold $L_{\rm cold,acc}$ is estimated to be $\sim 4.4\times 10^{34}$\,erg\,s$^{-1}$ which is shown by a dotted horizontal line in Figure {\ref{f10}}(i) and (ii). Interestingly, $L_{\rm cold,acc}$ is higher by a factor of about 4 than $L_{\rm cold}$ and using the more accurate threshold for accretion to proceed from a cold (non-ionized) disc, the estimated luminosities of the source lie below this threshold suggesting that accretion is likely mediated through a cold disc during the prolonged low luminosity period from 2018 December until 2024 February.
However, it seems that during the period 2005 April 6 (MJD 53466) until 2010 March 15 (MJD 55270.9), the pulsar was in a low luminosity state and likely not accreting from a cold disc for most of the time. Accretion from a cold disc has been suggested to occur during periods in between outbursts in GRO J1008-57 (lasting for about 200 d \citep{tsygankov2017stable}). In GX 304-1, this phenomenon has been observed in between Type I outbursts (lasting a few tens of days) and also after the cessation of regular outbursts (lasting for about 400 d \citep{escorial2018discovery}). The prolonged low-luminosity state (spanning a duration of about 5 yr as shown in Figure {\ref{f10}) powered by accretion from a cold disc in GX 304-1 might be the longest manifestation of this phenomenon reported in any accreting pulsar.}

\subsection{Accretion from stellar wind}
The companion stars in BeXRB systems contain massive stars and it is likely that accretion may proceed directly via the stellar wind. The companion star in GX 304-1 belongs to the B2V class \citep{mason1978optical,thomas1979lambda,parkes1980shell}
 which have typical terminal wind velocities ($v_w$) of $\sim \rm{800 ~km ~s^{-1}}$ and mass-loss rates of about $10^{-8} M_{\odot} \rm{~yr^{-1}}$ \citep{prinja1989ultraviolet,vink2000new}. Assuming that the gravitational potential energy of the captured stellar wind by the neutron star is entirely converted into X-rays, the X-ray luminosity ($L_{\rm wind}$) is given by \citep{reig2018discovery},

\begin{equation}
L_{\rm wind} \approx 4.7 \times 10^{37}
 M_{1.4}^3 \, R_6^{-1} \, M_{\rm *}^{-2/3} \, P_{\rm orb}^{-4/3} \,
 \dot{M}_{-6} \, v_{\rm w,8}^{-4} \, \,
  {\rm erg \, s^{-1}},
\end{equation}

where $M_{\rm *}$ is the mass of the companion star in solar masses,
$P_{\rm orb}$ is the orbital period of the binary in days while $\dot{M}$ and $v_w$ are expressed in units of $10^{-6} M_{\odot} \rm{~yr^{-1}}$ and $10^{-8} \rm{cm ~s^{-1}}$ respectively. Using $M_{\rm *}$=9.9 $M_{\odot}$ \citep{sugizaki2015luminosity}, $v_w \sim \rm{800 ~km ~s^{-1}}$, $\dot{M} \sim 10^{-8} M_{\odot} \rm{~yr^{-1}}$ for B2V type stars \citep{prinja1989ultraviolet,vink2000new}, $P_{\rm orb}\sim$132.2 d \citep{sugizaki2015luminosity}, $M=10 ~M_{\odot}$ and R=10 km, the estimated X-ray luminosity from the stellar wind is $\sim 4\times 10^{32}$\,erg\,s$^{-1}$. The estimated $L_{\rm wind}$ is lower by about a factor of 15-70 than the range of luminosities of the pulsar shown in Figure {\ref{f10}}(ii). The estimated X-ray luminosity during the low state from \textit{XMM-Newton} and \textit{NuStar} observations is $\sim 2.5\times 10^{33}$\,erg\,s$^{-1}$ in the 0.5-10 keV energy band and $\sim 3.6\times 10^{33}$\,erg\,s$^{-1}$ in the 3-20 keV energy band respectively, which are an order of magnitude higher than that can be powered solely by wind accretion. Thus, accretion from the stellar wind emanating from the Be star alone cannot explain the observed long-term low luminosity regime observed in this pulsar and it is possible that other favourable accretion mechanisms such as accretion through a cold (non-ionized) disc may operate simultaneously.

\subsection{Origin of soft excess in the spectra}
The hot BB excess ($kT_{\rm BB}$ >$\sim$ 1 keV) detected in GX 304-1 is similar to those detected in other BeXRBs RX J0440.9+4431 \citep{la2012xmm}, RX J0146.9+6121 \citep{la2006xmm}, 4U 0352+309
\citep{coburn2001discovery,la2007xmm}, RX J1037.5-5647 \citep{reig1999discovery,la2009xmm}, 3A 0535+262 \citep{mukherjee2005x}, 4U 2206+54 \citep{masetti2004look,torrejon2004evidence,reig2009discovery},
SAX J2103.5+4545 \citep{inam2004discovery},and SXP 1062 \citep{henault2012discovery,gonzalez2018multiwavelength}. These BeXRBs have long pulse periods (P > 100 s) and the hot BB excess were detected during low luminosity states ($\leq 5\times 10^{35}$\,erg\,s$^{-1}$) in these pulsars \citep{la2012xmm}. The spin period of GX 304-1 is about 275 s and the inferred luminosity during low states is $\sim 2.5\times 10^{33}$\,erg\,s$^{-1}$ and $\sim 3.6\times 10^{33}$\,erg\,s$^{-1}$ from  \textit{XMM-Newton} and \textit{NuStar} observations respectively, which suggests that GX 304-1 also belongs to this category of BeXRBs showing a soft X-ray excess.

It has been shown that in the low luminosity accretion powered pulsars ($L_{\rm X}\le10^{36}$ erg s$^{-1}$), the thermal component can be attributed to either emission by photo-ionized or collisionally heated diffuse gas or thermal emission from the surface of the neutron star \citep{hickox2004origin}. Assuming that the soft excess comes from the surface of the neutron star, the size of the emission region can be estimated using  R$_{col}\sim$ R$_{\rm NS}$ (R$_{\rm NS}$/R$_m$)$^{0.5}$ \citep{hickox2004origin}, where R$_{col}$ is the radius of the accretion column and R$_m$ is the magnetospheric radius. The magnetospheric radius is given by \citep{campana8m},

\begin{equation}
R_{\rm m} =
2\times 10^7\,{\dot M}_{15}^{-2/7}\,B_9^{4/7}\,
M_{1.4}^{-1/7}\,R_6^{12/7} ~{\rm cm},
\end{equation}

where ${\dot M}$ is the accretion rate (in units of $10^{15}\,{\rm \ g  \, s^{-1}}$), B, M and R  are the magnetic field, mass and radius of the neutron star respectively. Using $L_X \sim 5\times 10^{33}$\,erg\,s$^{-1}$, $M=1.4 ~M_{\odot}$ and R=10 km, the accretion rate is estimated to be $\sim 2.7 \times 10^{13}\,{\rm \ g  \, s^{-1}}$. Using $B\sim 4.7 \times 10^{12}$G \citep{yamamoto2011discovery}, the magnetospheric radius is estimated to $\sim 7 \times 10^{9}$cm. The estimated radius of the accretion column (which is an
estimate of the expected size of the polar cap) is R$_{col} \sim$ 120 m which is remarkably similar to the estimated black body emitting radius of about 100-110 m.
This suggests that the observed blackbody emission emanates from the polar cap of the neutron star.

\subsection{Exploring spectral energy distribution of the companion star}
\begin{figure}
\centering
  \includegraphics[width=\linewidth]{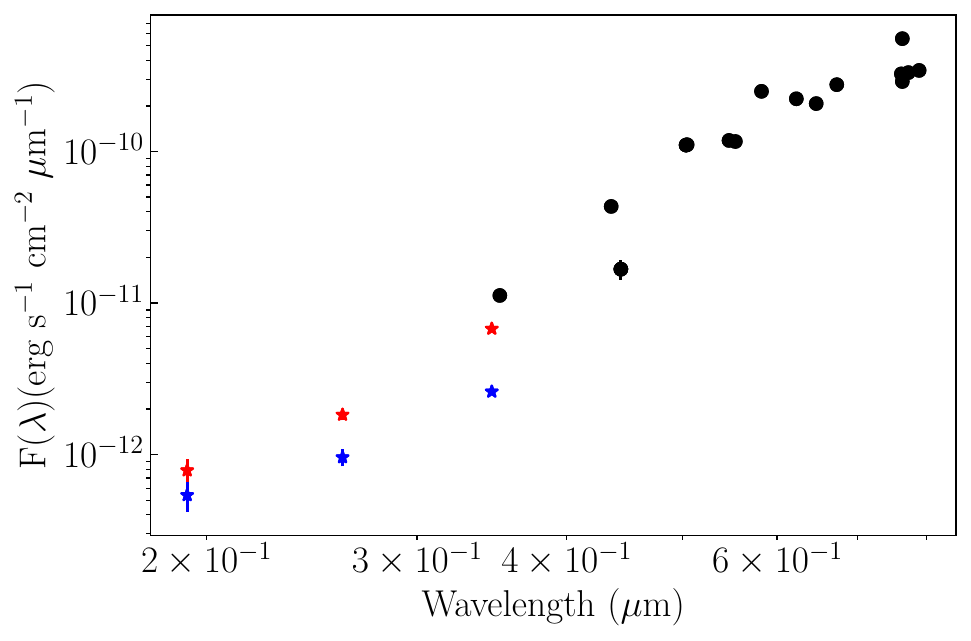}
  \caption{Spectral energy distribution for the companion star of GX 304-1. The black-filled circles show data taken from literature using the \textit{Gaia} \citep{prusti2016gaia,brown2021gaia} and \textit{TESS} \citep{stassun2019revised} observations and are obtained from the Vizier database \citep{ochsenbein2000vizier}. The blue and red stars show data from the \textit{Swift}/UVOT in U, UW1 and UW2 filters obtained during the quiescent state just before the regular X-ray outbursts began from the source (around MJD 55268) and that during the beginning of outbursts (around MJD 55550) respectively.}
 \label{f11}
\end{figure}

We searched the literature for photometric detection of the companion star at other wavelengths using the catalogues available in the VizieR database \citep{ochsenbein2000vizier}. The catalogue entries using the \textit{Gaia} \citep{prusti2016gaia,brown2021gaia} and \textit{TESS} \citep{stassun2019revised} observations were taken within $5''$ of the source position. The spectral energy distribution (SED) of the companion star is shown in Figure \ref{f11} plotted along with the \textit{Swift}/UVOT estimated fluxes in the U, UW1 and UW2 filters. We have shown the UVOT fluxes
obtained during the quiescent state just before the regular X-ray outbursts began from the source (around MJD 55268) and that during the beginning of outbursts (around MJD 55550). There is an indication of UV excess from the companion star when X-ray outbursts were detected from the neutron star compared to the time when the source was in a quiescent phase. The detailed SED modelling of the companion star is beyond the scope of this work.

\section{Conclusions}
We presented the timing and spectral studies of GX 304-1 using \textit{NuStar} and \textit{XMM-Newton} observations during low state. Pulsations are detected during the low state indicating that the low X-ray luminosity is powered by accretion onto the surface of the neutron star. We construct the long-term spin history of the pulsar and find that the pulsar switched from a long-term spin-up to a spin-down trend during a low luminosity state of the pulsar. The long-term photometric observations of the companion star suggest a possible decrease in its recent activity and lack of mass ejection events which trigger X-ray outbursts in this binary. The pulsar shows a prolonged low luminosity regime ($L_X \sim 10^{34}$\,erg\,s$^{-1}$) spanning nearly five years since 2018 December. The ultraviolet signatures of the companion star precedes the onset of X-ray outbursts in this source by about half a year and does not show any marked variation during the prolonged long-term quiescent state.
 We explore plausible mechanisms to explain the long-term spin-down and low luminosity manifestation in this pulsar. We detect soft X-ray excess in the spectra which can be attributed to thermal emission from the polar cap of the neutron star. Simultaneous X-ray, optical and ultraviolet monitoring observations of the binary are required to better understand the mechanisms of long-term spin-down, mass-loss dynamics from the companion star and accretion vagaries in this pulsar.

\begin{acknowledgement}
We are thankful to the reviewer for making constructive suggestions that improved the presentation of this paper.
This paper is dedicated to my mentor P. C. Agrawal who introduced me to the fascinating world of X-ray binaries.
ADC thanks Katja Pottschmidt for helpful suggestions regarding the \textit{NuStar} background. ADC thanks Bryan Irby for many prompt suggestions regarding an installation bug in the latest version of the \textsc{heasoft} package (ver 6.34). This research has made use of data from the \textit{NuStar} mission, a project led by the California Institute of Technology, managed by the Jet Propulsion Laboratory, and funded by the National Aeronautics and Space Administration. Data analysis was performed using the \textit{NuStar} Data Analysis Software (\textsc{nustardas}), jointly developed by the ASI Science Data Center (SSDC, Italy) and the California Institute of Technology (USA). This research is based on observations obtained with \textit{XMM-Newton}, an ESA science mission with instruments and contributions directly funded by ESA Member States and NASA. We acknowledge the use of public data from the \textit{Swift} data archive. This research has made use of software provided by the High Energy Astrophysics Science Archive Research Center (HEASARC), which is a service of the Astrophysics Science Division at NASA/GSFC and the High Energy Astrophysics Division of the Smithsonian Astrophysical Observatory.
This research has made use of the \textit{MAXI} \citep{matsuoka2009maxi} light curve provided by RIKEN, JAXA, and the \textit{MAXI} team. This research has also made use of the \textit{FERMI}/GBM \citep{meegan2009fermi} pulsar spin evolution history provided by the \textit{FERMI} team.
This research has made use of the \textit{Swift}/BAT \citep{krimm2013swift} light curve provided by the \textit{Swift} team. This research has made use of photometric observations of V850 Cen from ASAS-SN light curves \citep{shappee2014man,kochanek2017all}. We acknowledge with thanks the variable star observations from the AAVSO International Database contributed by observers worldwide and used in this research. This work has made use of data from the European Space Agency (ESA) mission
{\it Gaia} (\url{https://www.cosmos.esa.int/gaia}), processed by the {\it Gaia}
Data Processing and Analysis Consortium (DPAC,
\url{https://www.cosmos.esa.int/web/gaia/dpac/consortium}). Funding for the DPAC
has been provided by national institutions, in particular the institutions
participating in the {\it Gaia} Multilateral Agreement. This research has made use of the SIMBAD astronomical database \citep{wenger2000simbad} and NASA's Astrophysics Data System. This research has made use of Astropy \citep{robitaille2013astropy,price2018astropy}, Numpy \citep{harris2020array} and Matplotlib \citep{hunter2007matplotlib} packages.
ADC acknowledges support from ARIES through post-doctoral fellowship (AO/RA/2023/1197).
\end{acknowledgement}

\section*{Data availability}
The \textit{NuStar}, \textit{XMM-Newton} and \textit{Swift}/XRT data used in this research are available for download from the HEASARC (\url{https://heasarc.gsfc.nasa.gov/docs/archive.html}).
 The \textit{FERMI}/GBM pulsar spin evolution history for GX 304-1 provided by the \textit{FERMI} team is available at\\ \url{https://gammaray.nsstc.nasa.gov/gbm/science/pulsars/lightcurves/gx304m1.html}. The \textit{Swift}/BAT light curve is available at \url{https://swift.gsfc.nasa.gov/results/transients/weak/GX304-1/}. The \textit{MAXI} light curve is publicly available at \url{http://maxi.riken.jp/star_data/J1301-616/J1301-616.html}. The photometric observations of V850 Cen are available at \url{https://asas-sn.osu.edu/}.

\bibliography{example}

\clearpage
\newpage
\begin{appendix}

\clearpage
\onecolumn
\section{Additional tables}

\begin{longtable}{ccccccc}
\caption[{\it Swift}/XRT observations of GX 304-1]{{\it Swift}/XRT observations of GX 304-1 spanning the duration 2005 April 6 (MJD 53466) until 2010 March 15 (MJD 55270.9).} \label{tab1}\\
\hline\hline
S. no. & Obs Id & Date &  Exposure & XRT count rate (XCR) & $\frac{\rm{XCR}}{\rm{XCR}_{\rm{fixed}}}$$^{a}$ &  Luminosity$^{b}$  \\
 &  & MJD &  ks & $\rm{counts ~s^{-1}}$ &  &  $10^{34}$\,erg\,s$^{-1}$ \\
\hline

1	&	35072001	&	53466.01	&	3.50	&	0.552	$\pm$	0.021	&	4.18	&	5.20	$\pm$	0.2	\\
2	&	35072002	&	53998.05	&	4.14	&	1.229	$\pm$	0.023	&	9.31	&	11.57	$\pm$	0.22	\\
3	&	35072003	&	53999.05	&	8.02	&	1.093	$\pm$	0.016	&	8.28	&	10.29	$\pm$	0.15	\\
4	&	35072004	&	54091.85	&	4.56	&	0.885	$\pm$	0.019	&	6.70	&	8.33	$\pm$	0.18	\\
5	&	35072005	&	55263.23	&	1.14	&	1.19	$\pm$	0.06	&	9.02	&	11.20	$\pm$	0.56	\\
6	&	35072006	&	55263.69	&	1.16	&	0.537	$\pm$	0.024	&	4.07	&	5.06	$\pm$	0.23	\\
7	&	35072007	&	55264.23	&	1.17	&	0.541	$\pm$	0.024	&	4.10	&	5.09	$\pm$	0.23	\\
8	&	35072008	&	55264.70	&	1.16	&	0.546	$\pm$	0.024	&	4.14	&	5.14	$\pm$	0.23	\\
9	&	35072009	&	55265.25	&	0.88	&	0.85	$\pm$	0.03	&	6.44	&	8.00	$\pm$	0.28	\\
10	&	35072010	&	55265.78	&	1.00	&	0.593	$\pm$	0.027	&	4.49	&	5.58	$\pm$	0.25	\\
11	&	35072011	&	55266.24	&	1.40	&	0.543	$\pm$	0.025	&	4.11	&	5.11	$\pm$	0.24	\\
12	&	35072012	&	55266.78	&	1.28	&	0.392	$\pm$	0.02	&	2.97	&	3.69	$\pm$	0.19	\\
13	&	35072013	&	55267.12	&	0.85	&	0.47	$\pm$	0.026	&	3.56	&	4.42	$\pm$	0.24	\\
14	&	35072014	&	55267.78	&	1.27	&	0.551	$\pm$	0.023	&	4.17	&	5.19	$\pm$	0.22	\\
15	&	35072015	&	55268.25	&	1.42	&	0.63	$\pm$	0.023	&	4.77	&	5.93	$\pm$	0.22	\\
16	&	35072016	&	55268.78	&	1.27	&	0.528	$\pm$	0.022	&	4.00	&	4.97	$\pm$	0.21	\\
17	&	35072017	&	55269.18	&	1.32	&	0.491	$\pm$	0.021	&	3.72	&	4.62	$\pm$	0.2	\\
18	&	35072018	&	55269.93	&	0.98	&	0.52	$\pm$	0.026	&	3.94	&	4.90	$\pm$	0.24	\\
19	&	35072019	&	55270.39	&	1.21	&	0.64	$\pm$	0.025	&	4.85	&	6.02	$\pm$	0.24	\\
20	&	35072020	&	55270.86	&	1.27	&	0.613	$\pm$	0.024	&	4.64	&	5.77	$\pm$	0.23	\\

\hline

\end{longtable}

Notes. $^{a}$ XCR and $\rm{XCR_{fixed}}$=0.13 are the XRT count rates for the given epoch in the table and MJD 58272.3 (2018 June 3) respectively.\\
$^{b}$ Luminosity in the 0.5-100 keV energy range from \citet{tsygankov2019dramatic} using a distance of 2.01 kpc \citep{treuz2018distances} scaled using $\frac{\rm{XCR}}{\rm{XCR_{fixed}}}$.
\clearpage
\twocolumn

\clearpage
\onecolumn
\newpage
\begin{longtable}{ccccccc}
\caption[{\it Swift}/XRT observations of GX 304-1]{{\it Swift}/XRT observations of GX 304-1 spanning the period from 2018 December 17 (MJD 58469) until 2024 February 27 (MJD 60367.8).} \label{tab2}\\
S. no. & Obs Id & Date &  Exposure & XRT count rate (XCR) & $\frac{\rm{XCR}}{\rm{XCR}_{\rm{fixed}}}$$^{a}$ &  Luminosity$^{b}$  \\
 &  & MJD &  ks & $\rm{counts ~s^{-1}}$ &  &  $10^{34}$\,erg\,s$^{-1}$ \\
\hline

1       & 00035072138   &       58469.00        &       0.78    &       0.176   $\pm$   0.017   &       1.33    &       1.66    $\pm$   0.16    \\
2       & 00035072139   &       58483.41        &       0.66    &       0.209   $\pm$   0.02    &       1.58    &       1.97    $\pm$   0.19    \\
3       & 00035072140   &       58497.62        &       0.99    &       0.219   $\pm$   0.016   &       1.66    &       2.06    $\pm$   0.15    \\
4       & 00035072142   &       58514.96        &       0.98    &       0.151   $\pm$   0.015   &       1.14    &       1.42    $\pm$   0.14    \\
5       & 00035072144   &       58528.55        &       0.58    &       0.16    $\pm$   0.02    &       1.21    &       1.51    $\pm$   0.19    \\
6       & 00035072145   &       58539.38        &       0.91    &       0.116   $\pm$   0.015   &       0.88    &       1.09    $\pm$   0.14    \\
7       & 00035072146   &       58553.27        &       1.09    &       0.23    $\pm$   0.017   &       1.74    &       2.17    $\pm$   0.16    \\
8       & 00035072147   &       58567.42        &       1.04    &       0.139   $\pm$   0.013   &       1.05    &       1.31    $\pm$   0.12    \\
9       & 00035072148   &       58581.14        &       1.12    &       0.136   $\pm$   0.017   &       1.03    &       1.28    $\pm$   0.16    \\
10      & 00035072150   &       58643.04        &       0.90    &       0.171   $\pm$   0.015   &       1.30    &       1.61    $\pm$   0.14    \\
11      & 00035072151   &       58650.97        &       0.80    &       0.161   $\pm$   0.017   &       1.22    &       1.52    $\pm$   0.16    \\
12      & 00035072152   &       58657.92        &       0.90    &       0.118   $\pm$   0.014   &       0.89    &       1.11    $\pm$   0.13    \\
13      & 00035072153   &       58664.63        &       0.47    &       0.118   $\pm$   0.019   &       0.89    &       1.11    $\pm$   0.18    \\
14      & 00035072154   &       58671.14        &       0.91    &       0.125   $\pm$   0.014   &       0.95    &       1.18    $\pm$   0.13    \\
15      & 00035072155   &       58678.04        &       0.94    &       0.089   $\pm$   0.012   &       0.67    &       0.84    $\pm$   0.01    \\
16      & 00035072156   &       58685.08        &       0.92    &       0.162   $\pm$   0.015   &       1.23    &       1.52    $\pm$   0.14    \\
17      & 00035072157   &       58691.51        &       0.81    &       0.194   $\pm$   0.022   &       1.47    &       1.83    $\pm$   0.21    \\
18      & 00035072158   &       58700.16        &       0.76    &       0.277   $\pm$   0.022   &       2.10    &       2.61    $\pm$   0.21    \\
19      & 00035072159   &       58706.47        &       1.01    &       0.209   $\pm$   0.016   &       1.58    &       1.97    $\pm$   0.15    \\
20      & 00035072161   &       58845.96        &       0.91    &       0.215   $\pm$   0.017   &       1.63    &       2.02    $\pm$   0.16    \\
21      & 00035072162   &       58855.47        &       0.61    &       0.177   $\pm$   0.02    &       1.34    &       1.67    $\pm$   0.19    \\
22      & 00035072163   &       58860.24        &       0.99    &       0.104   $\pm$   0.015   &       0.79    &       0.98    $\pm$   0.01    \\
23      & 00035072164   &       58865.95        &       0.86    &       0.148   $\pm$   0.016   &       1.12    &       1.39    $\pm$   0.15    \\
24      & 00035072166   &       58875.98        &       0.60    &       0.194   $\pm$   0.022   &       1.47    &       1.83    $\pm$   0.21    \\
25      & 00035072167   &       58880.03        &       0.33    &       0.085   $\pm$   0.029   &       0.64    &       0.08    $\pm$   0.03    \\
26      & 00035072168   &       58885.22        &       0.67    &       0.153   $\pm$   0.021   &       1.16    &       1.44    $\pm$   0.2     \\
27      & 00035072169   &       59091.29        &       4.78    &       0.139   $\pm$   0.006   &       1.05    &       1.31    $\pm$   0.06    \\
28      & 00035072170   &       59436.08        &       2.64    &       0.12    $\pm$   0.009   &       0.91    &       1.13    $\pm$   0.08    \\
29      & 00035072171   &       60135.96        &       4.60    &       0.14    $\pm$   0.006   &       1.06    &       1.32    $\pm$   0.06    \\
30      & 00035072172   &       60188.77        &       1.09    &       0.163   $\pm$   0.026   &       1.23    &       1.53    $\pm$   0.24    \\
31      & 00035072173   &       60195.64        &       0.76    &       0.134   $\pm$   0.015   &       1.02    &       1.26    $\pm$   0.14    \\
32      & 00035072174   &       60202.44        &       0.90    &       0.164   $\pm$   0.015   &       1.24    &       1.54    $\pm$   0.14    \\
33      & 00035072175   &       60209.11        &       0.94    &       0.122   $\pm$   0.014   &       0.92    &       1.15    $\pm$   0.13    \\
34      & 00035072176   &       60216.11        &       1.01    &       0.18    $\pm$   0.024   &       1.36    &       1.69    $\pm$   0.23    \\
35      & 00035072177   &       60223.97        &       0.59    &       0.227   $\pm$   0.028   &       1.72    &       2.14    $\pm$   0.26    \\
36      & 00035072178   &       60230.17        &       0.94    &       0.103   $\pm$   0.013   &       0.78    &       0.97    $\pm$   0.01    \\
37      & 00035072179   &       60244.51        &       0.93    &       0.308   $\pm$   0.026   &       2.33    &       2.9     $\pm$   0.24    \\
38      & 00035072180   &       60251.84        &       0.86    &       0.064   $\pm$   0.011   &       0.48    &       0.6     $\pm$   0.01    \\
39      & 00035072181   &       60313.84        &       0.78    &       0.115   $\pm$   0.015   &       0.87    &       1.08    $\pm$   0.14    \\
40      & 00035072184   &       60331.45        &       0.91    &       0.182   $\pm$   0.017   &       1.38    &       1.71    $\pm$   0.16    \\
41      & 00035072186   &       60347.88        &       0.92    &       0.253   $\pm$   0.019   &       1.92    &       2.38    $\pm$   0.18    \\
42      & 00035072187   &       60349.99        &       0.97    &       0.132   $\pm$   0.019   &       1.00    &       1.24    $\pm$   0.18    \\
43      & 00035072188   &       60355.13        &       0.71    &       0.143   $\pm$   0.017   &       1.08    &       1.35    $\pm$   0.16    \\
44      & 00035072189   &       60361.38        &       1.08    &       0.128   $\pm$   0.014   &       0.97    &       1.2     $\pm$   0.13    \\
45      & 00035072190   &       60367.84        &       0.86    &       0.145   $\pm$   0.015   &       1.10    &       1.36    $\pm$   0.14    \\

\hline

\end{longtable}

Notes. $^{a}$ XCR and $\rm{XCR_{fixed}}$=0.13 are the XRT count rates for the given epoch in the table and MJD 58272.3 (2018 June 3) respectively.\\
$^{b}$ Luminosity in the 0.5-100 keV energy range from \citet{tsygankov2019dramatic} using a distance of 2.01 kpc \citep{treuz2018distances} scaled using $\frac{\rm{XCR}}{\rm{XCR_{fixed}}}$.
\clearpage
\twocolumn
\end{appendix}

\end{document}